\documentclass[preprint,tightenlines,aps,floats,nofootinbib]{revtex4}
  
\usepackage{epsf}

\newcommand{\bra}{\langle}
\newcommand{\ket}{\rangle}

\newcommand{\notE}{\ \hbox{{$E$}\kern-.60em\hbox{/}}}
\newcommand{\notp}{\ \hbox{{$p$}\kern-.43em\hbox{/}}}
\def\D0{\mbox{D\O}}

\preprint{\font\fortssbx=cmssbx10 scaled \magstep2
\hbox to \hsize{
\hskip1.2in %\raise.1in
\hbox{\fortssbx The University of Oklahoma}
\hskip0.2in $\vcenter{
                      \hbox{\bf OKHEP-05-01}
                      \hbox{\bf hep-ph/yymmnnn}
                      \hbox{October, 2007}}$ }
}
  
\begin{document}  

%\draft  
%-----------------------------------  
% Title  
%-----------------------------------  
\title{\vspace*{0.7in}
 SUSY QCD Corrections to Higgs Pair Production \\
 from Bottom Quark Fusion}
  
%-----------------------------------
% Authors
%-----------------------------------
\author{
Sally Dawson$^a$, Chung Kao$^{b}$, Yili Wang$^b$
}

%-----------------------------------
% Address
%-----------------------------------
\affiliation{
$^a$Department of Physics, Brookhaven National Laboratory, 
Upton, NY 11973, USA \\
$^b$Homer L. Dodge Department of Physics and Astronomy, 
University of Oklahoma \\ 
Norman, Oklahoma 73019, USA 
\vspace*{.5in}}

%\date{\today}
\thispagestyle{empty}

%-----------------------------------
%   Abstract
%-----------------------------------
\begin{abstract}

We present a complete next-to-leading order (NLO) calculation for 
the total cross section for inclusive Higgs pair production 
via bottom-quark fusion at the CERN Large Hadron Collider (LHC) in the minimal supersymmetric standard model (MSSM) and the  minimal supergravity 
model (mSUGRA). We emphasize the contributions of squark and gluino loops (SQCD)
and the decoupling properties of our results for heavy squark
and gluino masses. The enhanced couplings of the $b$ quark to
the Higgs bosons in supersymmetric models with large $\tan\beta$ yield 
 large NLO SQCD corrections in some regions of parameter space.
\end{abstract}

\pacs{PACS numbers: 14.80.Bn, 12.38.Bx, 13.85.Lg, 13.87.Ce}
%
%-----------------------------------------------------------------------
% Make Title Page
%-----------------------------------------------------------------------
\maketitle

%=======================================================================
% BEGIN MAIN TEXT
%=======================================================================
\newpage

%=======================================================================
%   BEGIN MAIN TEXT
%=======================================================================
%-----------------------------------------------------------------------
% 1 Introduction
%-----------------------------------------------------------------------
\section{Introduction}

In the standard model (SM), only one Higgs doublet is introduced and one neutral Higgs boson 
remains after electroweak symmetry breaking.
In the Minimal Supersymmetric Standard Model (MSSM)~\cite{Nilles:1983ge}, two Higgs doublets 
are required to break the electroweak symmetry.
The two Higgs doublets, $\phi_1$ and $\phi_2$, couple to fermions with weak isospin 
$-1/2$ and $+1/2$ respectively \cite{Gunion:1989we}. After spontaneous symmetry breaking, 
there remain five physical Higgs bosons: a
 singly charged Higgs boson $H^{\pm}$, two neutral CP-even scalars $h$  and $H$, 
and a neutral CP-odd pseudoscalar $A$. The Higgs potential is constrained by supersymmetry  
such that all tree-level Higgs boson masses and couplings are determined by just two 
independent parameters, commonly chosen to be the mass of the CP-odd pseudoscalar 
($M_A$) and the ratio of vacuum expectation values of the neutral Higgs fields 
($\tan\beta \equiv v_2/v_1$).

In the standard model, gluon fusion is the dominant process for producing a pair of 
Higgs bosons via triangle and box diagrams with internal top quarks 
and bottom quarks~\cite{Dicus:1987ic,Glover:1987nx,Plehn:1996wb,Dawson:1998py,Belyaev:1999mx,BarrientosBendezu:2001di,Binoth:2006ym}. 
 The Yukawa coupling of the Higgs boson to bottom quarks in the standard model is 
proportional to $m_b/v_{SM}$, where $v_{SM}$ is the vacuum-expectation value of the Higgs field,
 and 
is hence very weak.  Thus, at the Fermilab Tevatron and CERN Large Hadron Collider(LHC), 
the rate for inclusive Higgs pair production  from $b$ quark fusion is  small~\cite{Dawson:2006dm,BarrientosBendezu:2001di}.
However, it can  become significant in supersymmetric
models with large $\tan\beta$ since the Yukawa couplings of the Higgs bosons to the $b$ quark are enhanced 
by $1/\cos\beta$.
In the Minimal Supersymmetric Model (MSSM),  the production rate of pairs of the  lightest neutral Higgs  boson ($hh$)
from $b {\overline b}$ fusion is larger than the rate from the gluon-gluon initial state for
$\tan\beta \le 35$ and moderate values of $M_A$ ($M_A\sim 300~GeV$),
while the rate for pair production of the heavier neutral Higgs bosons ($HH$ and $AA$) from $b
{\overline b}$ fusion is highly suppressed relative to the gluon-gluon production mechanism at small tan$\beta$\cite{BarrientosBendezu:2001di,Jin:2005gw}. This makes it of interest to evaluate higher order corrections to the rates.

One of the most powerful ways to distinguish between 
the  Higgs boson of the standard model and those of   the 
MSSM is to measure the trilinear neutral Higgs boson couplings.  This
measurement is  extremely  challenging 
in most scenarios~\cite{Boudjema:1995cb,Djouadi:1999rc,Muhlleitner:2003me,Baur:2003gp,Moretti:2004wa}. 
The enhanced Higgs pair production rates from bottom quark fusion when $\tan\beta$ is 
large  provide discovery potential for determining  the trilinear Higgs couplings in the 
MSSM. The rate for Higgs boson production from bottom quark fusion\cite{Maltoni:2003pn}
 has been computed to 
NNLO\cite{Harlander:2003ai} and the electroweak 
and SUSY QCD corrections included to one-loop\cite{Dittmaier:2006cz}.  
Also, the rate for Higgs production in
association with a $b$ quark has been computed to 
NLO\cite{Campbell:2004pu,Dawson:2005vi,Dawson:2004wq,Dawson:2003kb,Dawson:2004sh,Dittmaier:2003ej}
 including SUSY QCD corrections\cite{Dawson:2007ur}. 

In this paper, we present a complete next-to-leading order (NLO) SUSY QCD
calculation for Higgs pair production via bottom quark fusion. The leading order (LO) 
process  is $b {\overline b}\to \phi\phi$ (where $\phi=h,H,A$)
 and the NLO cross section includes both ${\cal O}(\alpha_s)$ and ${\cal O}(1/\Lambda)$ corrections, 
where $\Lambda \equiv 
\ln( M_\phi/m_b )$~\cite{Olness:1987:ae,Barnett:1987jw,Stelzer:1998ni,Dicus:1998hs}. 
The subprocess $bg\to b \phi\phi$ contributes  ${\cal O}(1/\Lambda)$ corrections to the leading 
order  cross section for $b {\overline b}\to \phi\phi$. Theoretical predictions depend on 
the number of $b$ quarks tagged. Here, we consider only  inclusive processes in which there
are no tagged $b$ quarks.
The NLO pure QCD corrections have been computed in Ref.~\cite{Dawson:2006dm}.  The focus of this work is the
inclusion of the  ${\cal O}(\alpha_s)$ SUSY QCD (SQCD) corrections, which consist of
 squark and gluino loop contributions.  We also present a detailed 
study of the effects of the SUSY parameters on the production rate in the MSSM and mSUGRA
models.

In section II, we review the leading-order cross section for 
$pp \to \phi\phi$  from  $b\bar{b}$ fusion. 
In section III, we provide the complete  next-to-leading order (NLO) SQCD corrections
from gluino and squark loops
 for  $b\bar{b} \to \phi \phi$ production. 
Numerical results are given in Section IV and 
conclusions are drawn in Section V. In addition, there are three Appendices. 
Appendix A defines the  scalar integrals~\cite{'tHooft:1972fi,'tHooft:1978xw,van Oldenborgh:1989wn,van Oldenborgh:1990yc} used in 
our computation, Appendix B contains the coefficients used in computing the 
virtual SQCD corrections and Appendix C presents the Yukawa couplings, Higgs trilinear couplings and Higgs-squark-squark couplings in the MSSM.

%------------------------------------------------
% 2 Leading-order Cross Section for $b\bar{b} \to hh$
%------------------------------------------------
\section{Lowest Order Production in $ b \bar{b} \rightarrow \phi\phi $ }

\begin{figure}[htb]
\centering\leavevmode
\epsfxsize=3.5in
\epsfbox{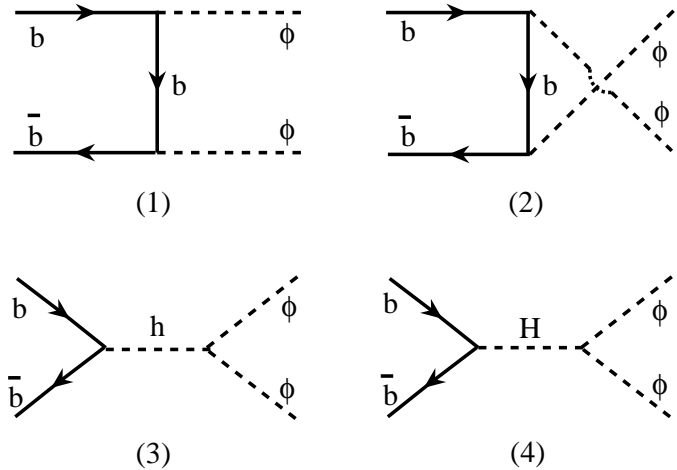}
\caption[]{The lowest order Feynman diagrams in the 
MSSM for $b \bar{b} \rightarrow \phi\phi$, with $\phi=h,H,A$.}
\label{fig:tree}
\end{figure}

The leading order (LO) inclusive cross section for  $pp \rightarrow \phi\phi$ via  $b\bar{b} \rightarrow \phi\phi $ is
\begin{eqnarray}
\sigma_{LO} = \int dx_1 dx_2 
              \left[ b(x_1) \bar{b}(x_2) +\bar{b}(x_1) b(x_2) \right]
              \hat{\sigma}_{LO}(s,t,u)(b\bar{b} \to \phi\phi)
\label{sigma:lo}
\end{eqnarray}
where $b(x)$ and $\bar{b}(x)$ are the LO parton distribution functions for 
bottom quarks in  the proton, $\hat{\sigma}_{LO}(s,t,u)$ is the parton
level cross section for $b\bar{b} \to \phi\phi $ and $s,t,u$ are the
Mandelstam variables. Fig.~\ref{fig:tree} shows the tree level Feynman 
diagrams for $b\bar{b} \to \phi\phi $ in the MSSM.

We assign momenta to the initial and the final state partons with 
\begin{equation}
b (p_1)\bar{b}(p_2) \rightarrow \phi(p_3) \phi(p_4) \, ,
\end{equation}
and $p_1 + p_2 = p_3 + p_4$. 
The $\phi b\bar{b}$ Yukawa coupling, $g_{\phi bb}$, and the $h\phi\phi$ 
and $H\phi\phi$ trilinear couplings,
  $g_{h\phi\phi}$  and   $g_{H\phi\phi}$, can be found in Appendix C. 
The lowest order rate can be expressed in terms of spinor structures:
\begin{eqnarray}
T_{1L}& = &\bar{v}(p_2) P_L u(p_1) \nonumber \\
T_{1R}& = &\bar{v}(p_2) P_R u(p_1)\, ,
\label{sdef}
\end{eqnarray}
with $P_{R,L}=(1\pm\gamma_5)/2$.
Following the simplified 
ACOT prescription~\cite{Aivazis:1993pi,Collins:1998rz,Kramer:2000hn}, 
we take $m_b = 0$ everywhere except in the Yukawa couplings. 
The tree level amplitudes of the $s-, t-$ and $u-$ channels are:
\begin{eqnarray}
M_s^0 
&= & -\left(\frac{g_{hbb} g_{h\phi\phi}}{s - M^2_h + i M_h \Gamma_h} 
+ \frac{g_{Hbb} g_{H\phi\phi}}{s - M^2_H + i M_H \Gamma_H}\right)\left(T_{1L} + T_{1R}\right) \delta_{\alpha\beta} \nonumber \\
&\equiv & \biggl(X_s^0 T_{1L}  + X_s^0 T_{1R}\biggr) \delta_{\alpha\beta} \nonumber \\
M_t^{0} &=& g_{\phi bb}^2\frac{1}{t}(\bar{v}(p_2) {\not p}_3
 P_L u(p_1) + \bar{v}(p_2) {\not p}_3 P_R u(p_1)) \delta_{\alpha\beta} \nonumber \\
 & \equiv & \biggl(\hat{M}^{0}_{tL} \delta_{ji}+\hat{M}^{0}_{tR}\biggr) \delta_{\alpha\beta} \nonumber \\
M_u^{0} &=&- g_{\phi bb}^2\frac{1}{u}(\bar{v}(p_2) {\not p}_3 P_L u(p_1) 
+ \bar{v}(p_2) {\not p}_3 P_R u(p_1)) \delta_{\alpha\beta} \nonumber \\
 & \equiv & \biggl(\hat{M}^{0}_{uL} \delta_{ji}+\hat{M}^{0}_{uR}\biggr) \delta_{\alpha\beta}\, , 
\label{smes}
\end{eqnarray}
where $\alpha, \beta$ are color indices.
The corresponding spin- and color-averaged  matrix elements squared,
 including interferences terms, are
\begin{eqnarray}
&&\bra |{M}_t^0|^2 \ket  =  \frac{g_{\phi bb}^4}{6} \left(\frac{u}{t} - \frac{M_h^4}{t^2}\right) \nonumber \\
&&\bra |{M}_u^0|^2 \ket  =  \frac{g_{\phi bb}^4}{6} \left(\frac{t}{u} - \frac{M_h^4}{u^2}\right) \nonumber \\
&& \bra Re({M}_t^0\bar{M}_u^{0}) \ket  =  - \frac{g_{\phi bb}^4}{6} \left(1 - \frac{M_h^4}{t u}\right) \nonumber\\
\bra |{M}_s^{0}|^2 \ket
&  = &\frac{g_{hbb}^2 g_{h\phi\phi}^2 s}{6 |s - M^2_h + i M_h \Gamma_h|^2}  
+ \frac{g_{Hbb}^2 g_{H\phi\phi}^2 s}{6 |s - M^2_H + i M_H \Gamma_H|^2}  \nonumber \\
&& + \frac{g_{hbb}g_{Hbb}g_{h\phi\phi} g_{H\phi\phi}}{3}\left(\frac{(s-M_h^2)(s-M_H^2) + M_h M_H \Gamma_h \Gamma_H}{|s - M^2_h + i M_h \Gamma_h|^2 \cdot |s - M^2_H + i M_H \Gamma_H|^2}\right) s \, . \nonumber \\
\end{eqnarray}
The LO element is then,
\begin{eqnarray}
M_0&=& M_t^0+M_u^0+M_s^0\nonumber \\
\bra |{M_0}|^2 \ket & = & \bra |{M_s^0}|^2 \ket + \bra |{M_t^0}|^2 \ket + \bra |{M_u^0}|^2 \ket +2 \bra Re({M_t}^0\bar{M_u^0}) \ket \, .\nonumber 
\end{eqnarray}
Finally, the parton level cross section for inclusive  $b\bar{b} \rightarrow \phi\phi $ production is
\begin{eqnarray}
\hat{\sigma}_{LO} & = &\int\frac{1}{2 s}\frac{1}{2}\bra |{M}_0|^2 \ket  d_2PS(b\bar{b} \rightarrow \phi\phi)\, , 
\end{eqnarray}
where $d_2PS(b\bar{b} \rightarrow \phi\phi)$ denotes the integral over the two-body phase space and the factor of $1/2$ is from the identical particles
in the final state.

\section{Next-To-Leading Order Corrections for $ b \bar{b} \rightarrow \phi\phi $}

The parton level NLO cross section is
\begin{eqnarray}
 \hat{\sigma}_{\rm NLO}(x_1,x_2, \mu) 
& = & \hat{\sigma}_{\rm LO}(x_1,x_2, \mu) 
     +\delta\hat{\sigma}_{NLO}(x_1, x_2, \mu) \nonumber \\
& \equiv & \hat{\sigma}_{\rm LO}(x_1,x_2, \mu) \left[ 1
     +{\hat\delta}_{\rm QCD}(x_1, x_2, \mu) + 
\hat{\delta}_{\rm SQCD}(x_1, x_2, \mu)\right]
\nonumber \\
  \delta\hat{\sigma}_{\rm NLO}(x_1,x_2, \mu) 
& \equiv & \delta\hat{\sigma}_{\rm QCD}(x_1, x_2, \mu)
     +\delta\hat{\sigma}_{\rm SQCD}(x_1, x_2, \mu)\, , 
\label{sigma:pnlo}
\end{eqnarray} 
where $\hat{\sigma}_{\rm LO}(x_1, x_2, \mu)$ is the leading order
(Born) cross section and $\delta\hat{\sigma}_{\rm NLO}(x_1, x_2, \mu)$ 
is the next-to-leading order correction to the Born cross section, 
$x_{1,2}$ are the  momentum fractions of the partons and $\mu = \mu_R$ is the
renormalization scale. The NLO correction $\delta \hat{\sigma}_{NLO}(x_1, x_2, \mu)$  
contains both the gluon QCD correction,  
$\delta \hat{\sigma}_{\rm QCD}(x_1, x_2, \mu)$,
 and the gluino-squark SQCD correction, 
 $\delta \hat{\sigma}_{\rm SQCD}(x_1, x_2, \mu)$. 
The gluon QCD correction includes the $O(\alpha_s)$ corrections, 
which contain virtual and real gluon emission contributions, 
as well as the $O(1/\Lambda)$ corrections from the $b g \to b \phi \phi$ subprocess. 
The gluino-squark SQCD correction contains $O(\alpha_s)$ gluino-sbottom loop contributions. 

Unlike down-type quarks, which only couple to  the down-type Higgs at tree level, 
the down-type squarks also couple to the up-type Higgs boson. 
This leads to mixing  in the sbottom mass matrix in the ${\tilde b}_L,
{\tilde b}_R$ basis, 
\begin{eqnarray}
m^2_{\tilde b_{1,2}} &=&
 \left( \begin{array}{cc}
m^2_{\tilde b_L} &  m_b (A_b - \mu \tan\beta) \\
m_b (A_b - \mu \tan\beta)  & m^2_{\tilde b_R} \\
\end{array} \right) \, ,
\label{bsmass}
\end{eqnarray}
where ${\tilde b_{1,2}}$ are the mass eigenstates.
The mass eigenstates $m_{\tilde b_1}$ and $m_{\tilde b_2}$ ($m_{\tilde b_1} \leq m_{\tilde b_2}$) are defined in terms of a mixing angle,
\begin{eqnarray}
 \left( \begin{array}{c}
\tilde{b}_1\\
\tilde{b}_2 \\
\end{array} \right)  &= & 
 \left( \begin{array}{cc}
\cos\theta_{\tilde b} &  \sin\theta_{\tilde b} \\
-\sin\theta_{\tilde b}  & \cos\theta_{\tilde b} \\
\end{array} \right)
 \left( \begin{array}{c}
\tilde{b}_L\\
\tilde{b}_R \\
\end{array} \right) \, .
\label{bsmixing}
\end{eqnarray}

\begin{figure}[htb]
\centering\leavevmode
\epsfxsize=5.5in
\epsfbox{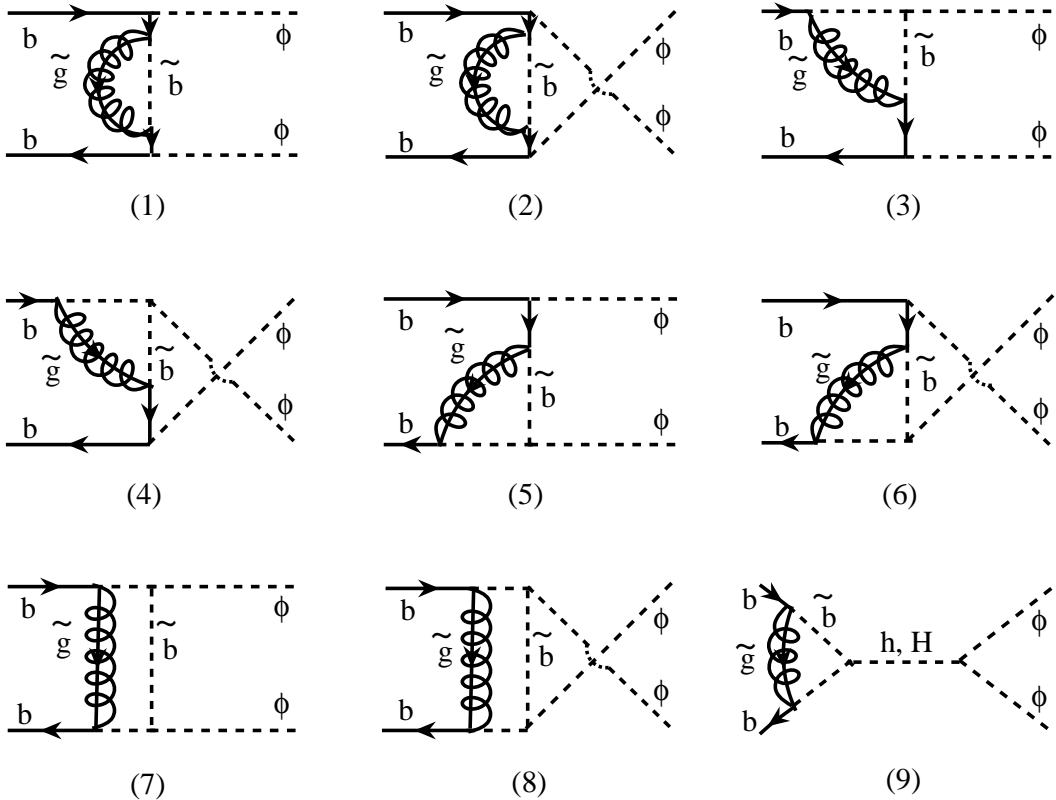}
\caption[]{One-loop SQCD virtual corrections to $b \bar{b} \rightarrow \phi\phi$.}
\label{fig:oneloop}
\end{figure}

The one-loop virtual gluino-sbottom Feynman diagrams for  
$b\bar{b} \rightarrow \phi\phi $ are shown in Fig.~\ref{fig:oneloop}. 
%------------------------------------------------
% FIG. 2
%------------------------------------------------
The amplitudes corresponding to each diagram in Fig.~\ref{fig:oneloop} are computed 
analytically and all tensor integrals are reduced to linear combinations of one-loop 
scalar functions. 
We sum all of the virtual amplitudes to obtain the contribution to the
 matrix element
 from gluino-sbottom loops in terms of the standard matrix elements 
of Eqs. \ref{sdef} and \ref{smes},
\begin{eqnarray}
M^{SQCD}_v &=& g_s^2 (T^a T^a)_{\alpha\beta}\left(\sum_{i=1,3,5,7}X_{i1} \hat{M}_{tL}^0 + \sum_{i=1,3,5,7} X_{i2} \hat{M}_{tR}^0 
+  \sum_{i=2,4,6,8} X_{i1} \hat{M}_{uL}^0 
\right. \nonumber \\
&&\left.+ \sum_{i=2,4,6,8} X_{i2} \hat{M}_{uR}^0 
+ \sum_{i=1}^{10} X_{i3} T_{1L} + \sum_{i=1}^{10} X_{i4} T_{1R}
\right) \, ,
\end{eqnarray}
where $i$ corresponds to the numbering of the diagrams in Fig.~\ref{fig:oneloop}.
 The analytic expressions for each coefficient $X$ can be found in appendix B. 
For simplification, we define:
\begin{eqnarray}
X_{tL} &= & \sum_{i=1,3,5,7} X_{i1} \,\quad\quad\quad\quad \, X_{tR} = \sum_{i=1,3,5,7} X_{i2} \nonumber \\
X_{uL} &= & \sum_{i=2,4,6,8} X_{i1}  \,\quad\quad\quad\quad \, X_{uR} = \sum_{i=2,4,6,8} X_{i2} \nonumber \\
X_{L} &=& \sum_{i=1}^{10} X_{i3}  \,\quad \quad \quad \quad \quad \quad \, X_{R} = \sum_{i=1}^{10} X_{i4}\, . 
\end{eqnarray}
The one-loop SQCD matrix element is then
\begin{eqnarray}
M_v^{SQCD}  &=& g_s^2 (T^a T^a)_{\alpha\beta}
 (X_{tR}\hat{M}_{tR}^0 + X_{tL} \hat{M}_{tL}^0 +X_{uR} \hat{M}_{uR}^0 + X_{uL}\hat{M}_{uL}^0 + X_{L} T_{1L} + X_{R} T_{1R} )\, , \nonumber \\
\label{equ:v}
\end{eqnarray}

Only ultraviolet (UV) divergences occur in the SUSY QCD corrections from the massive
gluino and squark loops.
These UV divergences are removed by the renormalization of the bottom quark wavefunction and
propagator and the  renormalization of  
the bottom quark mass in the Yukawa 
coupling~\cite{Braaten:1980yq,Beenakker:1993yr,Pierce:1997wu,Hafliger:2005aj,Berge:2007dz,Dawson:2007ur}.
We use the on-shell renormalization scheme of Ref. \cite{Berge:2007dz} and define the $b$ quark
self energy as,
\begin{eqnarray}
\Sigma^b(p)  &=&  \notp \left( \Sigma_V^b(p^2) 
-  \Sigma_A^b(p^2) \gamma_5\right) + m_b \Sigma_S^b(p^2) \nonumber \\
\delta \Sigma^b(p)  &=&  \notp \left( \delta Z_V^b -  \delta Z_A^b \gamma_5\right) 
- m_b \delta Z_V^b-\delta m_b \, ,
\end{eqnarray}
which gives the renormalized propagator,
\begin{equation}
\Sigma^b_{ren}=(\notp -m_b)\biggl(\Sigma_V^b+\delta  Z_V^b\biggr)+m_b\biggl(
\Sigma_V^b+\Sigma_S^b-{\delta m_b\over m_b}\biggr)\,.
\end{equation}
The on-shell renormalization condition requires,
\begin{eqnarray}
\Sigma^b(\notp =m_b)&=& 0\nonumber \\
lim_{\notp \rightarrow m_b}{\Sigma^b(p)\over \notp -m_b}&=&0 \, .
\end{eqnarray}

 Computing the bottom quark self energy from gluino-squark loops
and  ignoring contributions suppressed by powers of $m_b$,  we find:
\begin{eqnarray}
\Sigma_V^b(p^2) &=& -\frac{\alpha_s}{3\pi} \left[ B_1(p,m_{\tilde g}, m_{\tilde b_1}) 
+ B_1(p,m_{\tilde g}, m_{\tilde b_2})\right] \nonumber \\
\Sigma_S^b(p^2) &=& -\frac{\alpha_s}{3\pi}\biggl({m_{\tilde g}\over m_b}\biggr)
 \sin2\theta_{\tilde b} \left[ B_0(p,m_{\tilde g}, m_{\tilde b_2}) 
- B_0(p,m_{\tilde g}, m_{\tilde b_1})\right]\, ,
\end{eqnarray}
which yields,
\begin{eqnarray}
\delta Z_V^b  &=&  -\Sigma_V^b |_{p^2=m_b^2} \nonumber \\
\frac{\delta m_b}{m_b} &=& \left(\Sigma_V + \Sigma_S\right) |_{p^2=m_b^2}\, .
\end{eqnarray}

 The coupling of the $b$ squark to the up-type Higgs doublet
induces a modification of the tree-level relation between the 
bottom quark mass and its Yukawa coupling. At large $\tan\beta$, 
we absorb this modification by redefining the bottom quark mass occuring  in the
Yukawa coupling,\cite{Hall:1993gn,Carena:1999py,Guasch:2003cv,Carena:2006ai,Dittmaier:2006cz}
\begin{equation}
m_b \rightarrow \frac{m_b}{1 + \Delta_b}\, , 
\end{equation} 
where 
\begin{eqnarray}
\Delta_b = \frac{2\alpha_s(\mu_R)}{3\pi} m_{\tilde g} \mu \tan\beta I(m_{\tilde b_1}, m_{\tilde b_2}, m_{\tilde g}) \quad .
\end{eqnarray}
and the auxiliary function is defined as,
\begin{eqnarray}
I(a,b,c) = -\frac{1}{(a^2-b^2)(b^2-c^2)(c^2-a^2)}(a^2 b^2\ln\frac{a^2}{b^2}+b^2 c^2\ln\frac{b^2}{c^2}+c^2 a^2\ln\frac{c^2}{a^2}) \quad .
\end{eqnarray}
The contributions  to the bottom Yukawa couplings
which are enhanced at large $\tan\beta$ can be included
to all orders by making 
the following replacements~\cite{Hall:1993gn,Carena:1999py,Guasch:2003cv,Carena:2006ai,Dittmaier:2006cz}:
\begin{eqnarray}
g_{hbb} & \rightarrow & g_{hbb} \frac{1 - \Delta_b/(\tan\beta \tan\alpha)}{1 + \Delta_b} \nonumber \\
g_{Hbb} & \rightarrow & g_{Hbb} \frac{1 + \Delta_b \tan\alpha /\tan\beta}{1 + \Delta_b} \nonumber \\
g_{Abb} & \rightarrow & g_{Abb} \frac{1 - \Delta_b/\tan^2\beta}{1 + \Delta_b} \nonumber 
\, .
\end{eqnarray}
We use these effective Yukawa couplings in our NLO computation. 
To avoid double-counting, 
we add additional mass counterterms,
\begin{equation} 
\frac{\delta \tilde{m}_b^h}{m_b} = 
\Delta_b \left(1 + \frac{1}{\tan\alpha \tan\beta}\right),\quad 
\frac{\delta \tilde{m}_b^H}{m_b} = 
\Delta_b \left(1 - \frac{\tan\alpha}{\tan\beta}\right),\quad 
\frac{\delta \tilde{m}_b^A}{m_b} = 
\Delta_b \left(1 + \frac{1}{\tan^2\beta}\right),\quad
\end{equation} 
for $h$, $H$ and $A$ production, respectively.

\begin{figure}[htb]
\centering\leavevmode
\epsfxsize=3.5in
\epsfbox{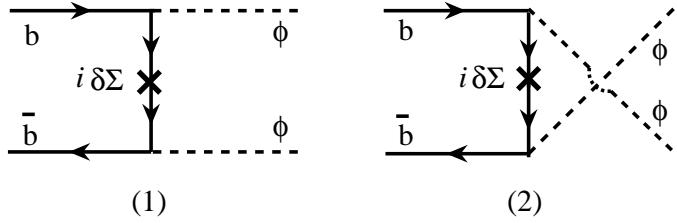}
\caption[]{$t-$ and $u-$ channel bottom quark propagator counterterm diagrams.}
\label{fg:counter}
\end{figure}

The counterterms arise from the $b$ quark wavefunction renormalization (a factor of $\delta
Z_b/2 $ for each external $b$ quark), the $b$ mass renormalization in the $\phi b{\overline b}$ 
couplings
(this occurs twice in the $t-$ and $u-$ channel diagrams, and once in the $s-$channel diagram),
and a factor of $-\delta Z_V$ from the mass renormalization on the internal $b$ quark
propagators shown in Fig. \ref{fg:counter}:
\begin{eqnarray}
M_{CT}& = &  
2\biggl(\frac{\delta m_b}{m_b} + \frac{\delta \tilde{m}^\phi_b}{m_b} \biggr) 
(M^0_t + M^0_u) 
+ \biggl( \delta Z_V +\frac{\delta m_b}{m_b} + \frac{\delta \tilde{m}^\phi_b}{m_b} \biggr) M^0_s
\, .
\end{eqnarray}
%------------------------------------------------
% FIG. 3
%------------------------------------------------
The  complete one-loop SUSY QCD contribution is,
\begin{eqnarray}
M^{SQCD} &=& M_0+ M_v^{SQCD} + M_{CT}\, ,
\end{eqnarray}
and is of course finite.
Finally, the SQCD contribution to the matrix element squared is,
\begin{eqnarray}
|M^{SQCD}|^2
& \equiv & \mid M_0\mid^2 +2 Re(  M^{SQCD}_v{\overline{M_0} }) 
+2 Re(  M_{CT}{\overline{M_{0}} }) \nonumber \\
& = & \mid M_0\mid^2+ 
2 Re(M_v^{SQCD} {\overline{M_0}})
+  4(\frac{\delta m_b}{m_b} + \frac{\delta \tilde{m}^\phi_b}{m_b} ) |M^0_t + M^0_u|^2 
\nonumber \\
&&+ 2(\delta Z_V + \frac{\delta m_b}{m_b} + \frac{\delta \tilde{m}^\phi_b}{m_b} ) |M^0_s|^2 \nonumber \\
\label{eq:v}
\end{eqnarray}
with the spin and color averaged result, 
\begin{eqnarray}
2\langle Re(M_{v}^{SQCD}{\overline{M_0}})\rangle
& = & \frac{4 \pi \alpha_s}{9}\left\{(X_{tR}  +  X_{tL}) |M_t^0|^2 + (X_{uR} +  X_{uL}) |M_u^0|^2 \right . \nonumber \\ 
& + &\left. ( X_{tR} +  X_{uR} +  X_{tL} + X_{uL}) Re(M_t^0 \overline M_u^0)\right. \nonumber \\ 
& +& \left .  2 Re[(X_L +X_R) X_s^{0*}] s \right\}\, . 
\end{eqnarray}

%------------------------------------------------
% 4 Results for Higgs pair production in bottom quark fusion
%------------------------------------------------
\section{RESULTS FOR HIGGS PAIR PRODUCTION FROM BOTTOM QUARK FUSION}

In this section, we present the next-to-leading-order inclusive cross 
sections for the production of a pair of neutral Higgs bosons via 
bottom quark fusion in the  MSSM and mSUGRA models at the CERN LHC.
 As in our previous paper~\cite{Dawson:2006dm}, we use the lowest order CTEQ6L1 parton distribution functions 
(PDFs)~\cite{Pumplin:2002vw} with a factorization scale $\mu_F$ 
and the leading-order evolution of the strong coupling
$\alpha_s(\mu_R)$ with a  renormalization scale $\mu_R$ 
to calculate the LO cross sections and use
the CTEQ6M PDFs with the next-to-leading-order evolution of 
$\alpha_s(\mu_R)$ to evaluate the NLO inclusive cross sections. 
For simplification, we use the same 
renormalization and factorization scales  $\mu_F = \mu_R$.
We evaluate the bottom quark mass occuring in the $\phi b {\overline b}$ Yukawa couplings 
using the ${\overline{MS}}$ mass, ${\overline m}_b(\mu)$, with a two-loop heavy
quark running mass with $m_b(pole)=4.7~GeV$ and the NLO evolution of the strong
coupling constant, modified to decouple the effects
of the top quark\cite{Marciano:1983pj,Nason:1987xz,Vermaseren:1997fq}.
The Higgs couplings are in Appendix C~\cite{Gunion:1989we,Haber:1984rc,Drees:2004,Tata:2006}
and we compute the Higgs boson masses to one-loop accuracy~\cite{Baer:1991yc}.

\subsection{Results in the  Minimal Supersymmetric Model (MSSM)}
%------------------------------------------------
% FIG. 
%------------------------------------------------

\begin{figure}[htb]
\centering\leavevmode
\epsfxsize=6.5in
\epsfbox{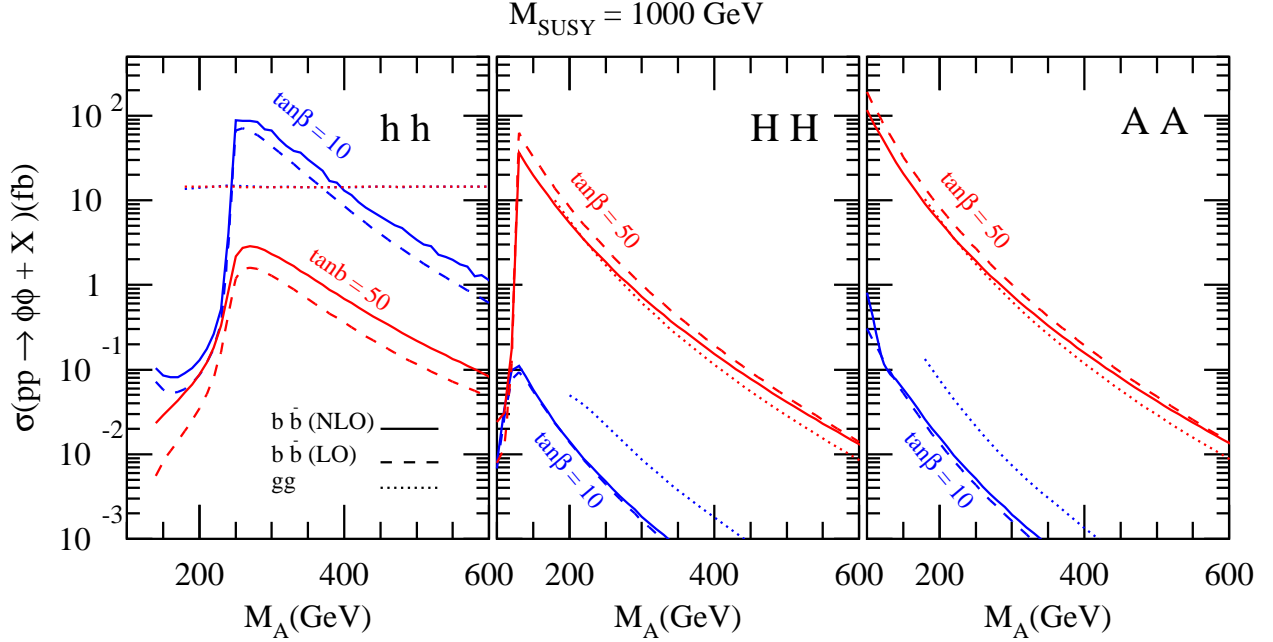}
\caption[]{Next-to-leading order cross sections 
$\sigma_{NLO}(pp \to \phi\phi +X)$ (solid) and leading order cross sections 
$\sigma_{LO}(pp \to \phi\phi +X)$ (dash) for Higgs pair production
from  bottom quark fusion versus 
the pseudoscalar Higgs mass ($M_A$) with $\sqrt{S}= 14$~TeV and
$\mu_R = \mu_F = M_{\phi}/2$.  The NLO cross sections include both
the pure QCD and the SQCD corrections.  
 We use $\tan\beta=50$ (red) and $\tan\beta=10$ (blue) with 
$M_{SUSY} = 1000$~GeV. Also shown are the cross sections of $g g$ fusion (dot). 
}
\label{fig:mssm1}
\end{figure}

 In Fig.~\ref{fig:mssm1}, we show the LO and NLO cross sections versus 
the pseudoscalar Higgs mass $M_A$. 
We assume $m_{\tilde g}=m_{\tilde{b_L}}=m_{\tilde{b_R}}=-A_b=\mu=M_{SUSY}$ and compute
the $b$ squark masses and mixing angles from Eqa. \ref{bsmass} and \ref{bsmixing}.
Our NLO cross sections include 
the ${\cal O}(\alpha_s)$ corrections from the 
$b {\overline b}$ initial state and the ${\cal O}(1/\Lambda)$ corrections 
from the $bg$ initial state, along with the SQCD corrections from gluino-sbottom loops.
  We show our results with $\tan\beta = 50$ (red) and $\tan\beta = 10$ (blue) 
at $M_{SUSY} = 1000$~GeV. To compare with $gg$ fusion, we also plot the cross section from the $gg$ initial state (dot)~\cite{Plehn:1996wb,BarrientosBendezu:2001di}.
We note that the 
cross sections for Higgs pair production in the MSSM are significantly
larger than in the standard model, due to enhancements at large $\tan\beta$.
The resonant enhancements due to $s-$ channel scalar exchange are clearly 
visible in the $b{\overline b}\rightarrow hh$ and $b {\overline b}
\rightarrow HH$ curves. At $\tan\beta = 10$, the cross section for pair of the lightest neutral Higgs  boson ($hh$) from $b {\overline b}$ fusion is much larger than the cross section from the gluon-gluon initial state, while the rate for pair production of the heavier neutral Higgs bosons ($HH$ and $AA$) from $b {\overline b}$ fusion is highly suppressed relative to the gluon-gluon production. But at $\tan\beta = 50$, gluon fusion dominates  pair of the lightest neutral Higgs  boson ($hh$) production and is comparible with $b {\overline b}$ production for the heavier neutral Higgs bosons ($HH$ and $AA$) production. 

%------------------------------------------------
% FIG. 
%------------------------------------------------

\begin{figure}[ht]
\centering\leavevmode
\epsfxsize=3.in
\epsfbox{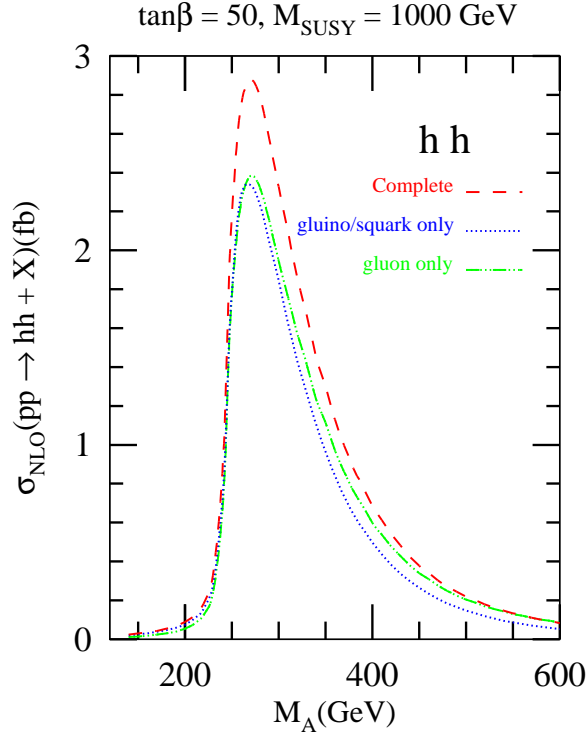}
\caption[]{Next-to-leading order cross section 
$\sigma_{NLO}(pp \to hh +X)$ in bottom quark fusion versus $M_A$ 
with $\sqrt{S}= 14$~TeV, $\tan\beta = 50$ and $M_{SUSY} = 1000$~GeV. 
 We plot the NLO cross section with only gluon QCD correction (dash-dot-dot, green), NLO cross section with only gluino-sbottom SQCD correction (dot, blue) and complete NLO cross section with both gluon QCD and gluino-sbottom SQCD corrections (dash, red).
}
\label{fig:mssm2}
\end{figure}

\begin{figure}[htb]
\centering\leavevmode
\epsfxsize=4in
\epsfbox{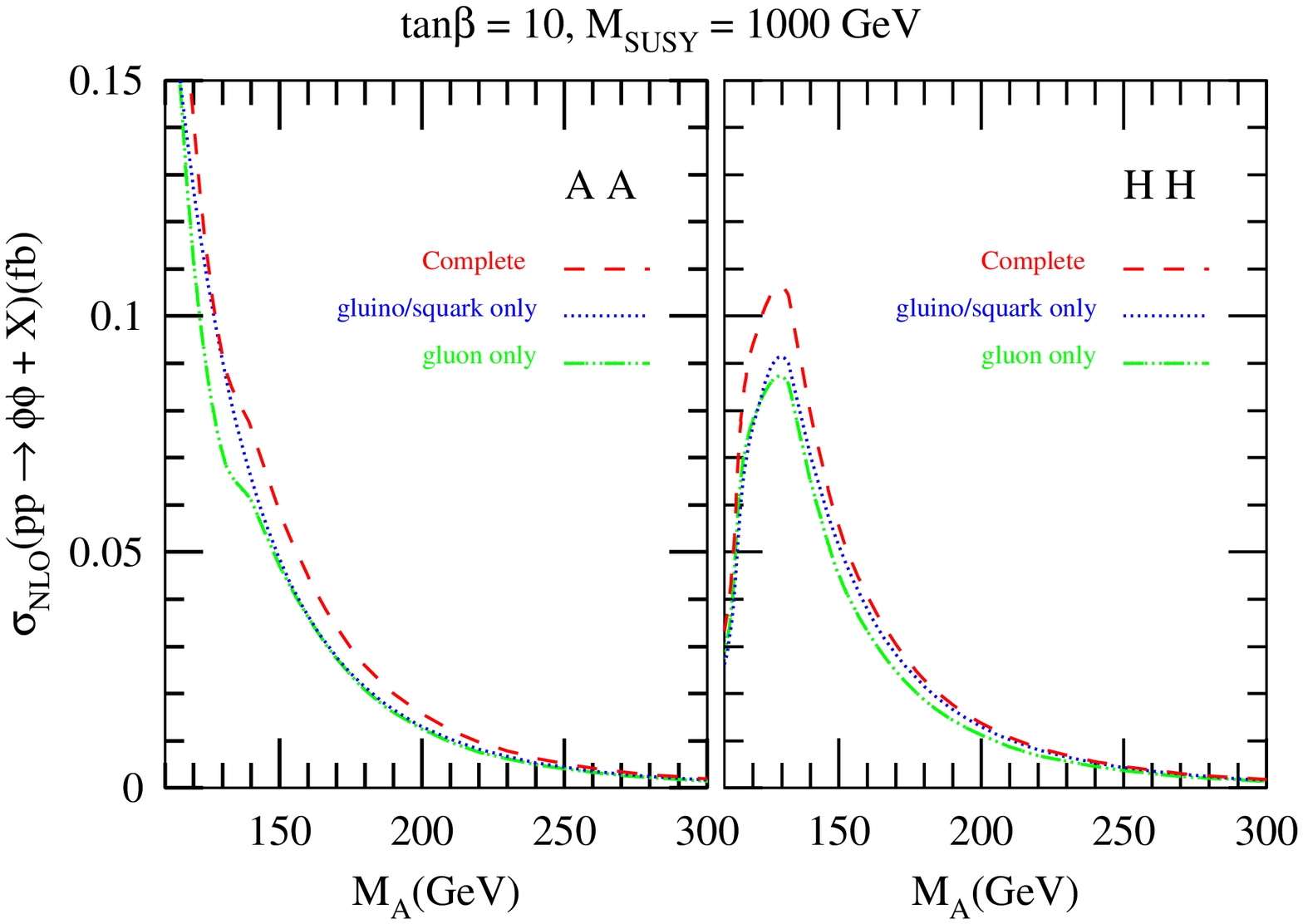}
\caption[]{Next-to-leading order cross sections  for (a)
$\sigma_{NLO}(pp \to HH +X)$ and (b) $\sigma_{NLO}(pp \to AA +X)$ from bottom quark fusion versus $M_A$ with $\sqrt{S}= 14$~TeV, $\tan\beta = 50$ and $M_{SUSY} = 1000$~GeV. 
We plot the
NLO cross section with only gluon QCD correction (dash-dot-dot, green), NLO cross section with only gluino-sbottom SQCD correction (dot, blue) and complete NLO cross section with both gluon QCD and gluino-sbottom SQCD corrections (dash, red).
}
\label{fig:mssm3}
\end{figure}

Figs.~\ref{fig:mssm2} and Fig.~\ref{fig:mssm3} show the NLO cross sections versus $M_A$ with $M_{SUSY} = 1000$~GeV and 
$\tan\beta = 50$. We present the  NLO cross section
 with only gluon QCD corrections (dash-dot-dot,green), NLO cross section with only gluino-sbottom SQCD corrections (dot,blue),and the complete NLO cross section with QCD and SQCD corrections together (dash, red). We note:
\begin{itemize}
\item The NLO SQCD correction to $b\bar{b} \to hh $ is small.
The dominant contribution to the
NLO correction to light Higgs pair production is  from the pure  QCD contribution.
\item  For heavy Higgs pair and pseudoscalar Higgs pair production,
the SUSY QCD corrections become dominant. The pure gluon NLO contribution
is much smaller in magnitude 
than the contribution from SQCD.
\end{itemize}

\begin{figure}[htb]
\centering\leavevmode
\epsfxsize=5in
\epsfbox{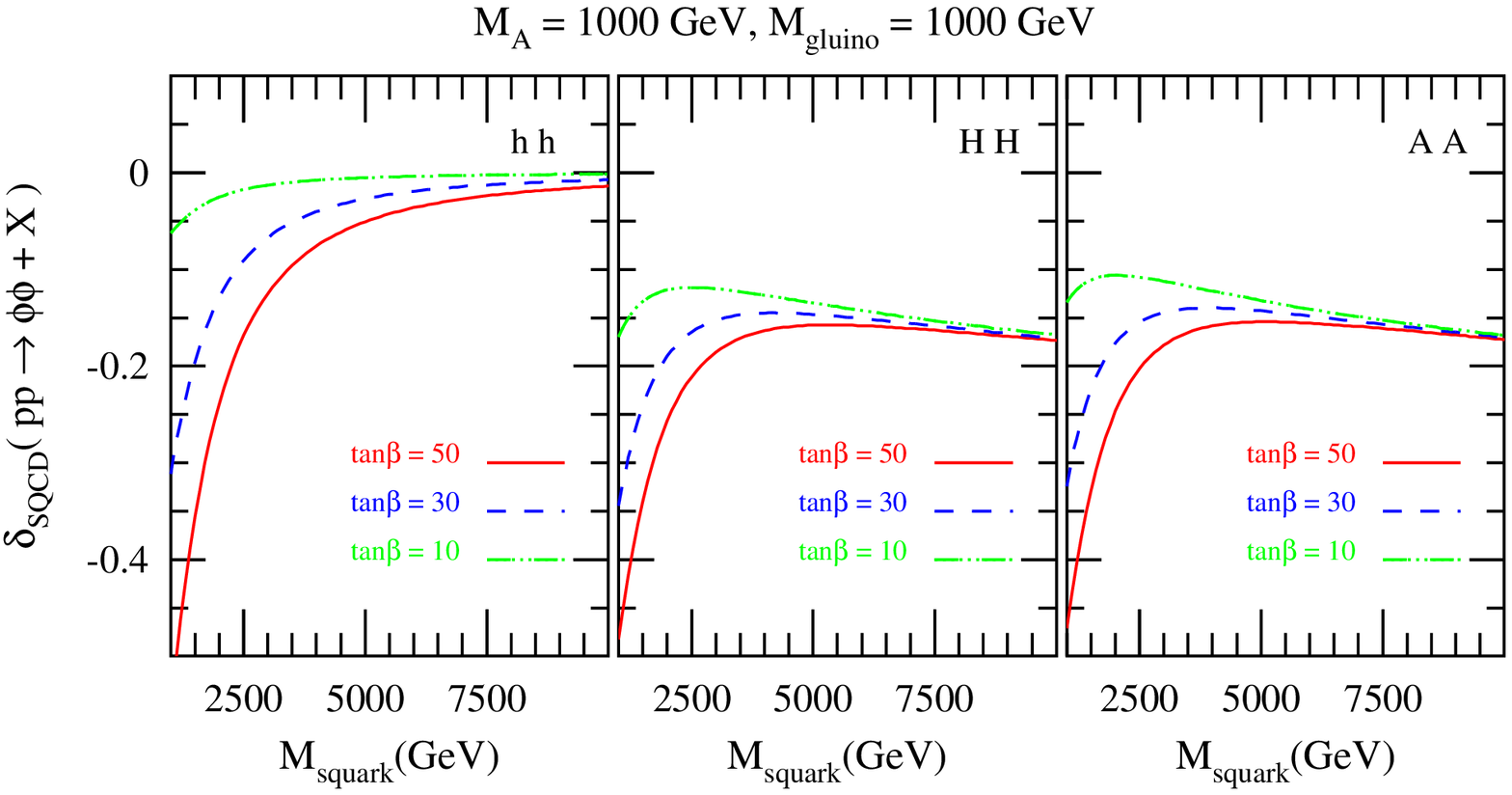}
\caption[]{ $\delta_{SQCD}$ versus squark mass $m_{\tilde q}$ with $M_A = m_{\tilde g} 
= 1000$~GeV and $\sqrt{S} = 14$~TeV for (a) $b\bar{b} \rightarrow hh $, (b) $b\bar{b} \rightarrow HH$ and (c) $b\bar{b} \rightarrow AA$. $\tan\beta = 10$ (dash-dot-dot, green),$\tan\beta = 30$ (dash, blue) and $\tan\beta = 50$ (solid, red). 
}
\label{fig:mssm4}
\end{figure}

\begin{figure}[htb]
\centering\leavevmode
\epsfxsize=5in
\epsfbox{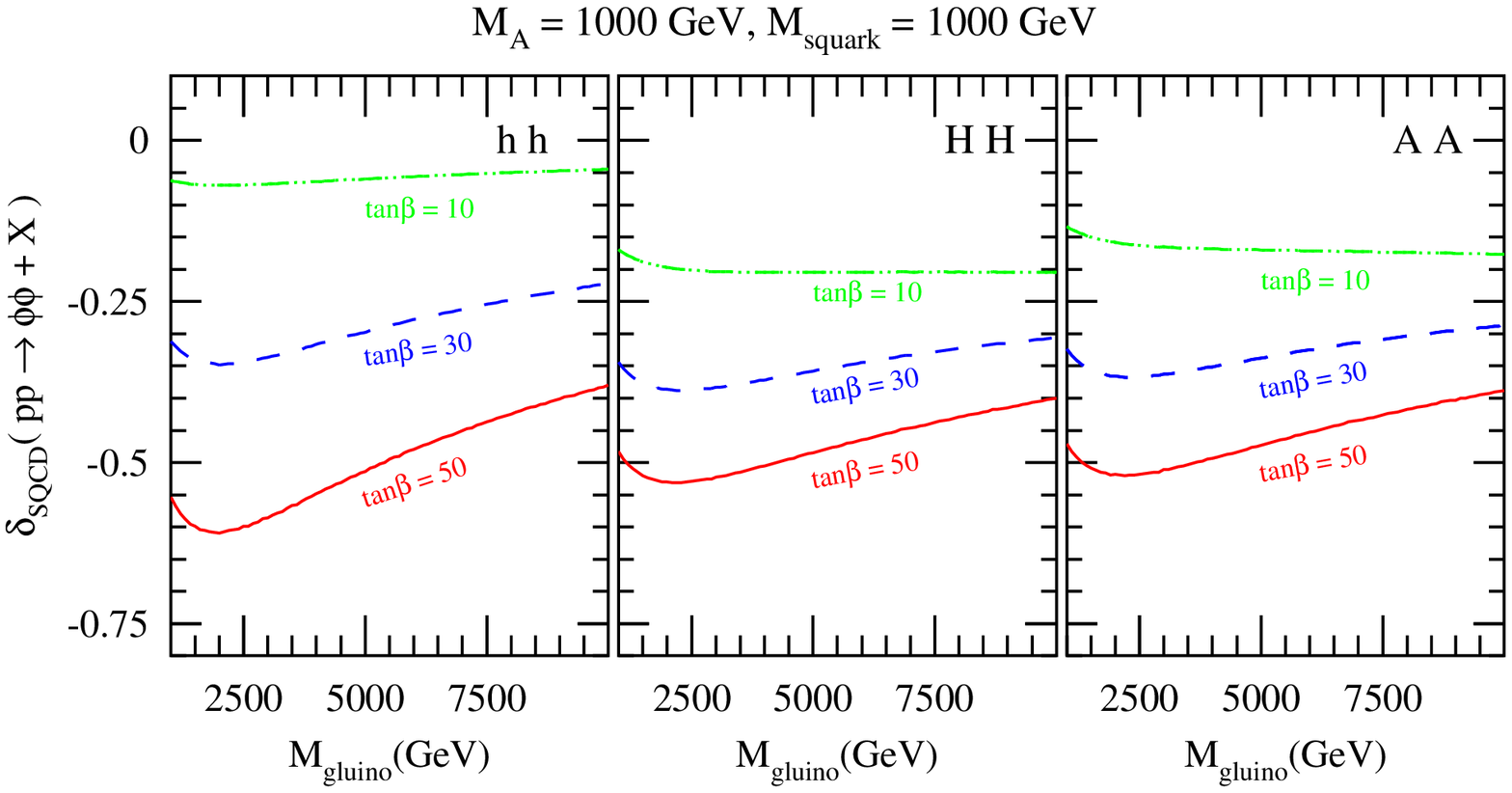}
\caption[]{ $\delta_{SQCD}$  versus gluino mass, $m_{\tilde g}$, with 
$M_A = m_{\tilde {b_{L,R}}}=\mu=1000$~GeV and $\sqrt{S} = 14$~TeV for (a) $b\bar{b} \rightarrow hh $, (b) $b\bar{b} \rightarrow HH$ and (c) $b\bar{b} \rightarrow AA$. $\tan\beta = 10$ (dash-dot-dot, green),$\tan\beta = 30$ (dash, blue) and $\tan\beta = 50$ (solid, red). 
  }
\label{fig:mssm5}
\end{figure}
In Fig.~\ref{fig:mssm4}, we plot the ratio of the NLO SQCD correction
normalized to the Born cross section, $\delta_{\rm SQCD}$,
 with $M_A =m_{\tilde g}=-A_b= 1000$~GeV, and $\tan\beta = 10$ (dash-dot-dot, green), 
$tan\beta = 30$ (dash, blue) and $tan\beta = 50$ (solid, red).  
In the limit of  large squark masses, $\delta_{SQCD}$  approaches a common 
non-zero constant for $HH$ and $AA$ production.  Light  Higgs pair production, however, 
decouples for large $M_A$ and large SUSY masses. This decoupling behaviour
is also seen in the decay $h\rightarrow b {\overline b}$~\cite{Haber:2000kq,Guasch:2003cv}
and the production process $bg\rightarrow bh$\cite{Dawson:2007ur}. 

In Fig.~\ref{fig:mssm5}, we fix $M_A$ and all squark masses to be $1000$~GeV  
and plot $\delta_{SQCD}$ versus the  gluino mass $m_{\tilde g}$ 
with  $tan\beta = 10$ (dash-dot-dot, green),  $tan\beta = 30$ (dash, blue) 
and $tan\beta = 50$ (solid, red).   This figure does not demonstrate a decoupling 
behaviour.   

\subsection{Results in the Minimal Supergravity Model (mSUGRA)}

In this model, supersymmetry is assumed to be broken in a hidden sector 
consisting of fields
that interact with the usual particles and their superpartners only via
gravity.  Supersymmetry breaking is communicated to the visible sector via
gravitational interactions. Within the mSUGRA framework, it is assumed that 
at some high scale (frequently taken to be $\sim M_{GUT}$) all scalar fields have a common SUSY breaking mass $M_0$, all gauginos have a common mass $M_{1/2}$, and all soft SUSY breaking scalar trilinear couplings have a common value $A_0$. 
Electroweak symmetry breaking is assumed to occur radiatively. This
fixes the magnitude of the superpotential parameter $\mu$. The soft SUSY
breaking bilinear Higgs boson mass parameter can be eliminated in favour
of $\tan\beta$, so that the model
is completely specified by the parameter set:
\begin{equation}
M_0,\ M_{1/2},\ A_0,\ \tan\beta,\ sign(\mu ) .
\end{equation}
All the sparticle masses and couplings required for phenomenological analysis can be obtained via renormalization group evolution between the scale of
grand unification and the weak scale.

\begin{figure}[]
\centering\leavevmode
\epsfxsize=6in
\epsfbox{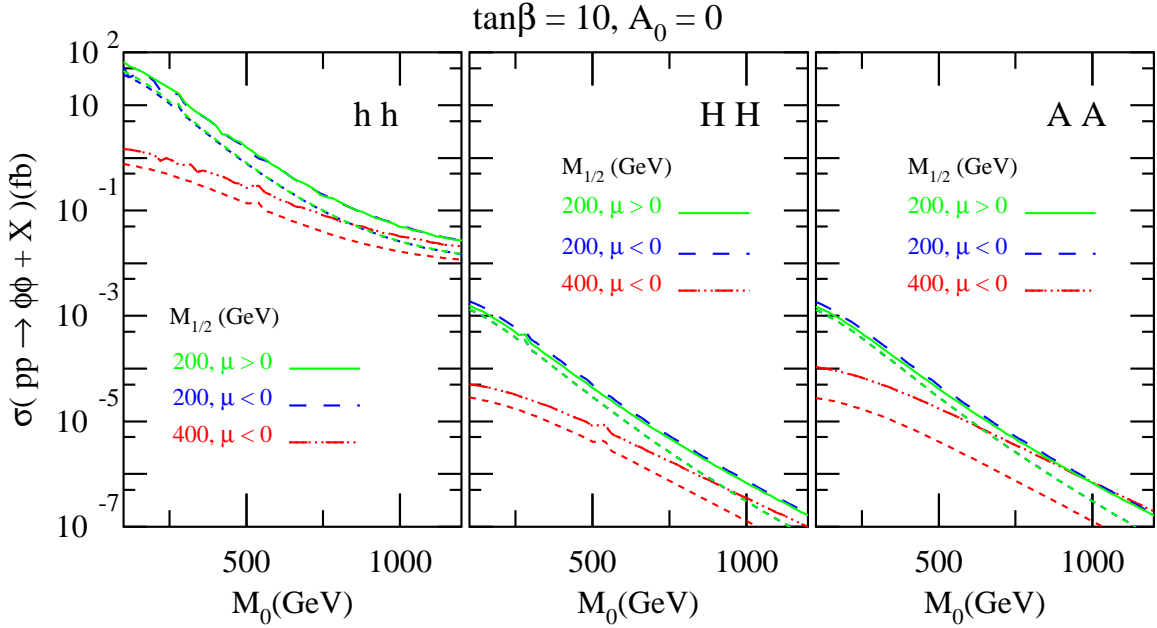}

\caption[]{
$\sigma_{NLO}$  versus  $M_0 $ with $\sqrt{S}= 14$~TeV and $\tan\beta = 10$.
Three cases are shown with $\mu > 0$ for $M_{1/2} = 200$ (blue)and $400$~GeV (red), $\mu < 0$ for $M_{1/2} = 200$~GeV(green). Also shown are the LO cross sections, $\sigma_{\rm LO}$ (dot).
  }
\label{fig:msugra1}
\end{figure} 
%------------------------------------------------
% FIG. 7
%------------------------------------------------

\begin{figure}[ht]
\centering\leavevmode
\epsfxsize=6in
\epsfbox{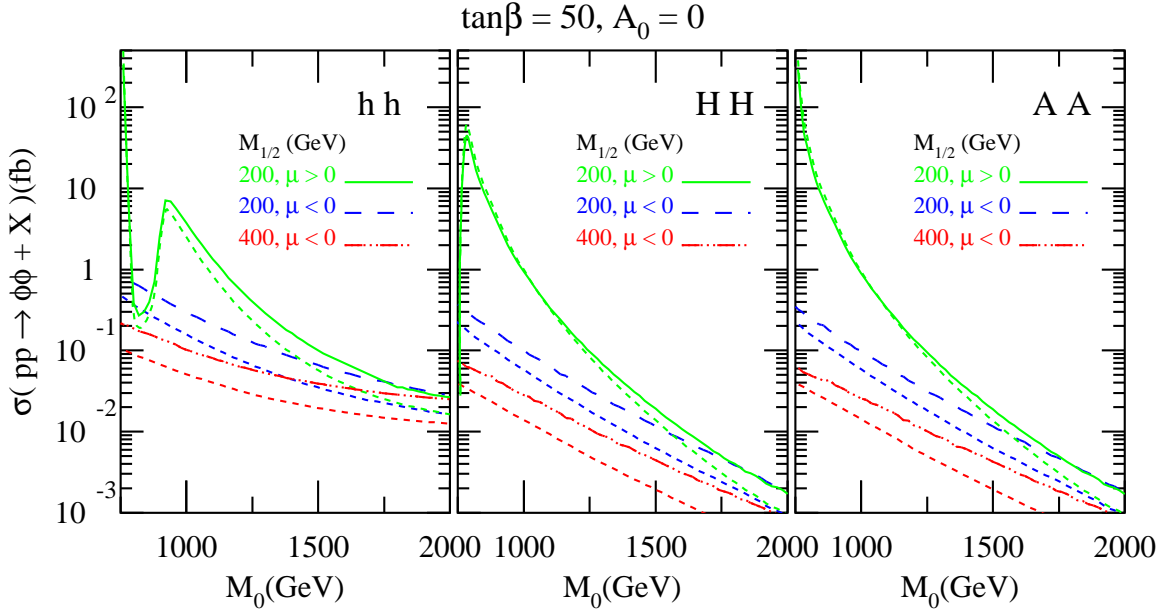}
\caption[]{
$\sigma_{NLO}$  versus  $M_0 $ with $\sqrt{S}= 14$~TeV and $\tan\beta = 50$.
Three cases are shown with $\mu > 0$ for $M_{1/2} = 200$ (blue)
and $400$~GeV (red), $\mu < 0$ for $M_{1/2} = 200$~GeV(green). Also shown are the LO cross sections,
 $\sigma_{\rm LO}$ (dot).
  }
\label{fig:msugra2}
\end{figure}

We show the LO (dot) and NLO (solid) cross sections versus $M_0$ with 
$A_0 = 0$ for $\tan\beta = 10$ in Fig.~\ref{fig:msugra1} and $\tan\beta = 50$ 
in Fig.~\ref{fig:msugra2}. We plot six curves in each frame, three LO 
cross sections (dot) and three NLO cross sections (solid). The NLO
curves include both the pure QCD and the SQCD contributions. As shown in the 
graphs, blue lines are cross sections with $M_{1/2} = 200$~GeV, $\mu < 0$, 
green lines have  $M_{1/2} = 200$~GeV, $\mu > 0$, and red lines are for 
$M_{1/2} = 400 $~GeV, $\mu < 0$. Comparing  Figs.~\ref{fig:msugra1} and
~\ref{fig:msugra2} we see a strong dependence on  $\tan\beta$.
\begin{itemize}
\item When $\tan\beta = 10$, flipping the sign of $\mu$ 
has little effect on either the LO or the
NLO cross sections with $M_{1/2} = 200$~GeV. 
As $\tan\beta$ increases to 50, we notice that flipping the sign of
 $\mu$ has a large effect
when $M_0 < 1200$~GeV as shown in Fig.~\ref{fig:msugra2}.
\item For large $M_0$, the cross sections approach 
a common value, independent of $M_{1/2}$ and $\mu$.  
\end{itemize}

%-------------------------------------------------------------------
% 4 Conclusions
%-----------------------------------------------------------------------
\section{Conclusions}

In this paper we presented the complete ${\cal O}(\alpha_s)$ 
 SUSY QCD corrections 
to neutral Higgs pair production from  bottom quark fusion in 
the Minimal Supersymmetric Model and the Minimal Supergravity Model 
at the CERN LHC. 
The effects of the SUSY QCD
 corrections from sbottom and gluino loops are significant in some
regions of parameter space and the decoupling of the SQCD effects is
only recovered for light Higgs pair production in the large $M_A$ limit.

%-----------------------------------------------------------------------
% Acknowledgments
%-----------------------------------------------------------------------
\section*{Acknowledgments}

We are grateful to Jianwei Qiu  and Chris Jackson for beneficial discussions.
This research was supported in part by the U.S. Department of Energy
under Grants No.~DE-AC02-76CH1-886,
No.~DE-FG02-04ER41305, No.~DE-FG02-03ER46040 and
DE-AC02-98CH10886.  SD thanks the SLAC  theory group for
their hospitality, where this
work was completed.

%----------------------------------------------------------------------
% Appendix
%-----------------------------------------------------------------------
\appendix
\section{
The scalar integrals}

\begin{eqnarray}
{i\over 16\pi^2} B_0(p, m_1, m_2) & = & \int {d^nk\over (2\pi)^n}
 \frac{1}{(k^2-m_1^2) [(k+p)^2-m_2^2]} \nonumber \\
{i\over 16\pi^2}\not{p} B_{11}(p,m_1,m_2) & = &
  \int{d^nk\over (2\pi)^n}  \frac{\not k}{(k^2-m_1^2) [(k+p)^2-m_2^2]} \nonumber \\
{i\over 16\pi^2}C_{0}(p_1, p_2, m_1, m_2,m_3) & = & 
 \int {d^nk\over (2\pi)^n} \frac{1}{(k^2-m_1^2) [(k+p_1)^2-m_2^2][(k+p_1+p_2)^2 -m_3^2]} \nonumber
\end{eqnarray}

\begin{eqnarray}
&&{i\over 16\pi^2}\biggl(\not p_1 C_{11}(p_1, p_2, m_1, m_2,m_3) + \not p_2  C_{12}(p_1, p_2, m_1, m_2,m_3)\biggr) \nonumber \\
&&= \int  {d^nk\over (2\pi)^n}\frac{\not k}{(k^2-m_1^2) [(k+p_1)^2-m_2^2][(k+p_1+p_2)^2 -m_3^2]} \nonumber 
\end{eqnarray}

\begin{eqnarray}
&&{i\over 16\pi^2}\biggl(\not p_1 D_{11}(p_1, p_2,p_3, m_1, m_2,m_3,m_4) + \not p_2  D_{12}(p_1, p_2,p_3, m_1, m_2,m_3,m_4)
\nonumber \\
&& + \not p_3  D_{13}(p_1, p_2, m_1, m_2,m_3,m_4)\biggr)\nonumber \\
&&=  \int {d^nk\over (2\pi)^n}\frac{\not k}{(k^2-m_1^2) [(k+p_1)^2-m_2^2][(k+p_1+p_2)^2 -m_3^2][(k+p_1+p_2+p_3)^2-m_4^2]} \nonumber \\
&&\nonumber \\
&&D_{123}(p_1, p_2,p_3, m_1, m_2,m_3,m_4) \nonumber \\
&&= D_{12}(p_1, p_2,p_3, m_1, m_2,m_3,m_4) - D_{13}(p_1, p_2,p_3, m_1, m_2,m_3,m_4) \nonumber 
\end{eqnarray}

\section{X Coefficients}
\subsection{X coefficients for $b \bar{b} \to h h$ and $b \bar{b} \rightarrow
H H $}
\begin{eqnarray}
X_{11} & = & \frac{1}{8\pi^2} \left[ B_1(-p_1 + p_3, m_{\tilde{g}}, m_{\tilde b_1}) \sin^2\theta_{\tilde b} + B_1(-p_1 + p_3, m_{\tilde{g}}, m_{\tilde b_2}) \cos^2\theta_{\tilde b}\right] \nonumber \\
X_{12} & = & \frac{1}{8\pi^2} \left[ B_1(-p_1+p_3, m_{\tilde{g}}, m_{\tilde b_1}) \cos^2\theta_{\tilde b} + B_1(-p_1+p_3, m_{\tilde{g}}, m_{\tilde b_2}) \sin^2\theta_{\tilde b}\right] \nonumber \\
X_{13} & = & -\frac{1}{8\pi^2} \frac{g_{\phi bb}^2}{t} m_{\tilde g} \left[B_{0}(-p_1+p_3, m_{\tilde g}, m_{\tilde b_1}) - B_{0}(-p_1+p_3, m_{\tilde g}, m_{\tilde b_2}) \right]\sin\theta_{\tilde b} \cos\theta_{\tilde b} \nonumber \\
X_{14} &=& X_{13} \nonumber \\
 \nonumber \\
X_{2i} &= & X_{1i}( t \leftrightarrow u, p_3 \leftrightarrow p_4) \quad i = 1,2,3,4 \nonumber \\
 \nonumber \\
X_{31} &= & \frac{1}{8\pi^2}\frac{g_{\phi11}} {g_{\phi bb}} m_{\tilde g} C_0(-p_1,p_3, m_{\tilde g}, m_{\tilde b_1},m_{\tilde b_1}) \sin\theta_{\tilde b} \cos\theta_{\tilde b}\nonumber \\
 &+& \frac{1}{8\pi^2}\frac{g_{\phi12}} {g_{\phi bb}} m_{\tilde g} C_0(-p_1,p_3, m_{\tilde g}, m_{\tilde b_1},m_{\tilde b_2}) \cos^2\theta_{\tilde b}  \nonumber \\
&-& \frac{1}{8\pi^2}\frac{g_{\phi12}} {g_{\phi bb}} m_{\tilde g} C_0(-p_1,p_3, m_{\tilde g}, m_{\tilde b_2},m_{\tilde b_1}) \sin^2\theta_{\tilde b}  \nonumber \\
&-& \frac{1}{8\pi^2}\frac{g_{\phi22}} {g_{\phi bb}} m_{\tilde g} C_0(-p_1,p_3, m_{\tilde g}, m_{\tilde b_2},m_{\tilde b_2}) \sin\theta_{\tilde b} \cos\theta_{\tilde b} \nonumber \\
X_{32} & = &  \frac{1}{8\pi^2}\frac{g_{\phi11}} {g_{\phi bb}} m_{\tilde g} C_0(-p_1,p_3, m_{\tilde g}, m_{\tilde b_1},m_{\tilde b_1}) \sin\theta_{\tilde b} \cos\theta_{\tilde b}\nonumber \\
&-& \frac{1}{8\pi^2}\frac{g_{\phi12}} {g_{\phi bb}} m_{\tilde g} C_0(-p_1,p_3, m_{\tilde g}, m_{\tilde b_1},m_{\tilde b_2}) \sin^2\theta_{\tilde b}  \nonumber \\
&+& \frac{1}{8\pi^2}\frac{g_{\phi12}} {g_{\phi bb}} m_{\tilde g} C_0(-p_1,p_3, m_{\tilde g}, m_{\tilde b_2},m_{\tilde b_1}) \cos^2\theta_{\tilde b}  \nonumber \\
&-& \frac{1}{8\pi^2}\frac{g_{\phi22}} {g_{\phi bb}} m_{\tilde g} C_0(-p_1,p_3, m_{\tilde g}, m_{\tilde b_2},m_{\tilde b_2}) \sin\theta_{\tilde b} \cos\theta_{\tilde b} \nonumber \\
X_{33} &= & -\frac{1}{8\pi^2} g_{\phi bb} g_{\phi11} C_{12}(-p_1,p_3, m_{\tilde g}, m_{\tilde b_1},m_{\tilde b_1}) \cos^2\theta_{\tilde b} \nonumber \\
&+& \frac{1}{8\pi^2} g_{\phi bb} g_{\phi12} C_{12}(-p_1,p_3, m_{\tilde g}, m_{\tilde b_1},m_{\tilde b_2}) \sin\theta_{\tilde b} \cos\theta_{\tilde b}  \nonumber \\
&+& \frac{1}{8\pi^2} g_{\phi bb} g_{\phi12} C_{12}(-p_1,p_3, m_{\tilde g}, m_{\tilde b_2},m_{\tilde b_1}) \sin\theta_{\tilde b} \cos\theta_{\tilde b}  \nonumber \\
&-& \frac{1}{8\pi^2} g_{\phi bb} g_{\phi22} C_{12}(-p_1,p_3, m_{\tilde g}, m_{\tilde b_2},m_{\tilde b_2}) \sin^2\theta_{\tilde b} \nonumber \\
X_{34} &= & -\frac{1}{8\pi^2} g_{\phi bb} g_{\phi11} C_{12}(-p_1,p_3, m_{\tilde g}, m_{\tilde b_1},m_{\tilde b_1}) \sin^2\theta_{\tilde b} \nonumber \\
&-& \frac{1}{8\pi^2} g_{\phi bb} g_{\phi12} C_{12}(-p_1,p_3, m_{\tilde g}, m_{\tilde b_1},m_{\tilde b_2}) \sin\theta_{\tilde b} \cos\theta_{\tilde b}  \nonumber \\
&-& \frac{1}{8\pi^2} g_{\phi bb} g_{\phi12} C_{12}(-p_1,p_3, m_{\tilde g}, m_{\tilde b_2},m_{\tilde b_1}) \sin\theta_{\tilde b} \cos\theta_{\tilde b}  \nonumber \\
&-& \frac{1}{8\pi^2} g_{\phi bb} g_{\phi22} C_{12}(-p_1,p_3, m_{\tilde g}, m_{\tilde b_2},m_{\tilde b_2}) \cos^2\theta_{\tilde b} \nonumber \\
 \nonumber \\
X_{4i} &= & X_{3i}( t \leftrightarrow u, p_3 \leftrightarrow p_4) \quad i = 1,2,3,4
 \nonumber \\
 \nonumber \\
X_{51} &= & \frac{1}{8\pi^2}\frac{g_{\phi11}} {g_{\phi bb}} m_{\tilde g} C_0(p_2,-p_4, m_{\tilde g}, m_{\tilde b_1},m_{\tilde b_1})  \sin\theta_{\tilde b} \cos\theta_{\tilde b}\nonumber \\
& - & \frac{1}{8\pi^2}\frac{g_{\phi12}} {g_{\phi bb}} m_{\tilde g} C_0(p_2,-p_4, m_{\tilde g}, m_{\tilde b_2},m_{\tilde b_1}) \sin^2\theta_{\tilde b}  \nonumber \\
& + & \frac{1}{8\pi^2}\frac{g_{\phi12}} {g_{\phi bb}} m_{\tilde g} C_0(p_2,-p_4, m_{\tilde g}, m_{\tilde b_1},m_{\tilde b_2}) \cos^2\theta_{\tilde b}  \nonumber \\
& - & \frac{1}{8\pi^2}\frac{g_{\phi22}} {g_{\phi bb}} m_{\tilde g} C_0(p_2,-p_4, m_{\tilde g}, m_{\tilde b_2},m_{\tilde b_2}) \sin\theta_{\tilde b} \cos\theta_{\tilde b}\nonumber \\
X_{52} & = & \frac{1}{8\pi^2}\frac{g_{\phi11}} {g_{\phi bb}} m_{\tilde g} C_0(p_2,-p_4, m_{\tilde g}, m_{\tilde b_1},m_{\tilde b_1})  \sin\theta_{\tilde b} \cos\theta_{\tilde b}\nonumber \\
&+& \frac{1}{8\pi^2}\frac{g_{\phi12}} {g_{\phi bb}} m_{\tilde g} C_0(p_2,-p_4, m_{\tilde g}, m_{\tilde b_2},m_{\tilde b_1}) \cos^2\theta_{\tilde b}  \nonumber \\
&-& \frac{1}{8\pi^2}\frac{g_{\phi12}} {g_{\phi bb}} m_{\tilde g} C_0(p_2,-p_4, m_{\tilde g}, m_{\tilde b_1},m_{\tilde b_2}) \sin^2\theta_{\tilde b}  \nonumber \\
&-& \frac{1}{8\pi^2}\frac{g_{\phi22}} {g_{\phi bb}} m_{\tilde g} C_0(p_2,-p_4, m_{\tilde g}, m_{\tilde b_2},m_{\tilde b_2}) \sin\theta_{\tilde b} \cos\theta_{\tilde b}\nonumber \\
X_{53} &= & -\frac{1}{8\pi^2} g_{\phi bb} g_{\phi11} C_{12}(p_2,-p_4, m_{\tilde g}, m_{\tilde b_1},m_{\tilde b_1}) \sin^2\theta_{\tilde b} \nonumber \\
&-& \frac{1}{8\pi^2} g_{\phi bb} g_{\phi12} C_{12}(p_2,-p_4, m_{\tilde g}, m_{\tilde b_2},m_{\tilde b_1}) \sin\theta_{\tilde b} \cos\theta_{\tilde b}  \nonumber \\
&-& \frac{1}{8\pi^2} g_{\phi bb} g_{\phi12} C_{12}(p_2,-p_4, m_{\tilde g}, m_{\tilde b_1},m_{\tilde b_2}) \sin\theta_{\tilde b} \cos\theta_{\tilde b}  \nonumber \\
&-& \frac{1}{8\pi^2} g_{\phi bb} g_{\phi22} C_{12}(p_2,-p_4, m_{\tilde g}, m_{\tilde b_2},m_{\tilde b_2}) \cos^2\theta_{\tilde b} \nonumber \\
X_{54} &= & -\frac{1}{8\pi^2} g_{\phi bb} g_{\phi11} C_{12}(p_2,-p_4, m_{\tilde g}, m_{\tilde b_1},m_{\tilde b_1}) \cos^2\theta_{\tilde b} \nonumber \\
&+& \frac{1}{8\pi^2} g_{\phi bb} g_{\phi12} C_{12}(p_2,-p_4, m_{\tilde g}, m_{\tilde b_2},m_{\tilde b_1}) \sin\theta_{\tilde b} \cos\theta_{\tilde b}  \nonumber \\
&+& \frac{1}{8\pi^2} g_{\phi bb} g_{\phi12} C_{12}(p_2,-p_4, m_{\tilde g}, m_{\tilde b_1},m_{\tilde b_2}) \sin\theta_{\tilde b} \cos\theta_{\tilde b}  \nonumber \\
&-& \frac{1}{8\pi^2} g_{\phi bb} g_{\phi22} C_{12}(p_2,-p_4, m_{\tilde g}, m_{\tilde b_2},m_{\tilde b_2}) \sin^2\theta_{\tilde b} \nonumber \\
 \nonumber \\
X_{6i} &= & X_{5i}( t \leftrightarrow u, p_3 \leftrightarrow p_4) \quad i = 1,2,3,4
 \nonumber \\ \nonumber \\
X_{71} &=& \frac{1}{8\pi^2} \frac{g_{\phi11}^2}{g_{\phi bb}^2} t  D_{123}(-p_1,p_3,p_4, m_{\tilde g}, m_{\tilde b_1},m_{\tilde b_1},m_{\tilde b_1}) \cos^2\theta_{\tilde b} \nonumber \\
&-& \frac{1}{8\pi^2} \frac{g_{\phi11} g_{\phi12}}{g_{\phi bb}^2} t D_{123}(-p_1,p_3,p_4, m_{\tilde g}, m_{\tilde b_1},m_{\tilde b_1},m_{\tilde b_2}) \sin\theta_{\tilde b} \cos\theta_{\tilde b} \nonumber \\
&+& \frac{1}{8\pi^2} \frac{g_{\phi12}^2}{g_{\phi bb}^2} t D_{123}(-p_1,p_3,p_4, m_{\tilde g}, m_{\tilde b_1},m_{\tilde b_2},m_{\tilde b_1}) \cos^2\theta_{\tilde b}\nonumber \\
&-& \frac{1}{8\pi^2} \frac{g_{\phi22} g_{\phi12}}{g_{\phi bb}^2} t D_{123}(-p_1,p_3,p_4, m_{\tilde g}, m_{\tilde b_1},m_{\tilde b_2},m_{\tilde b_2}) \sin\theta_{\tilde b} \cos\theta_{\tilde b} \nonumber \\
&-& \frac{1}{8\pi^2} \frac{g_{\phi11} g_{\phi12}}{g_{\phi bb}^2} t D_{123}(-p_1,p_3,p_4, m_{\tilde g}, m_{\tilde b_2},m_{\tilde b_1},m_{\tilde b_1}) \sin\theta_{\tilde b} \cos\theta_{\tilde b} \nonumber \\
& +& \frac{1}{8\pi^2} \frac{g_{\phi12}^2}{g_{\phi bb}^2} t D_{123}(-p_1,p_3,p_4, m_{\tilde g}, m_{\tilde b_2},m_{\tilde b_1},m_{\tilde b_2}) \sin^2\theta_{\tilde b} \nonumber \\
& -& \frac{1}{8\pi^2} \frac{g_{\phi22} g_{\phi12}}{g_{\phi bb}^2} t D_{123}(-p_1,p_3,p_4, m_{\tilde g}, m_{\tilde b_2},m_{\tilde b_2},m_{\tilde b_1}) \sin\theta_{\tilde b} \cos\theta_{\tilde b}\nonumber \\
&+& \frac{1}{8\pi^2} \frac{g_{\phi22}^2}{g_{\phi bb}^2} t  D_{123}(-p_1,p_3,p_4, m_{\tilde g}, m_{\tilde b_2},m_{\tilde b_2},m_{\tilde b_2}) \sin^2\theta_{\tilde b}\nonumber \\
X_{72} &=& \frac{1}{8\pi^2} \frac{g_{\phi11}^2}{g_{\phi bb}^2} t D_{123}(-p_1,p_3,p_4, m_{\tilde g}, m_{\tilde b_1},m_{\tilde b_1},m_{\tilde b_1}) \sin^2\theta_{\tilde b} \nonumber \\
&+& \frac{1}{8\pi^2} \frac{g_{\phi11} g_{\phi12}}{g_{\phi bb}^2} t D_{123}(-p_1,p_3,p_4, m_{\tilde g}, m_{\tilde b_1},m_{\tilde b_1},m_{\tilde b_2}) \sin\theta_{\tilde b} \cos\theta_{\tilde b}\nonumber \\
&+& \frac{1}{8\pi^2} \frac{g_{\phi12}^2}{g_{\phi bb}^2} t D_{123}(-p_1,p_3,p_4, m_{\tilde g}, m_{\tilde b_1},m_{\tilde b_2},m_{\tilde b_1}) \sin^2\theta_{\tilde b}\nonumber \\
&+& \frac{1}{8\pi^2} \frac{g_{\phi22} g_{\phi12}}{g_{\phi bb}^2} t D_{123}(-p_1,p_3,p_4, m_{\tilde g}, m_{\tilde b_1},m_{\tilde b_2},m_{\tilde b_2}) \sin\theta_{\tilde b} \cos\theta_{\tilde b} \nonumber \\
&+& \frac{1}{8\pi^2} \frac{g_{\phi11} g_{\phi12}}{g_{\phi bb}^2} t D_{123}(-p_1,p_3,p_4, m_{\tilde g}, m_{\tilde b_2},m_{\tilde b_1},m_{\tilde b_1}) \sin\theta_{\tilde b} \cos\theta_{\tilde b} \nonumber \\
&+& \frac{1}{8\pi^2} \frac{g_{\phi12}^2}{g_{\phi bb}^2} t D_{123}(-p_1,p_3,p_4, m_{\tilde g}, m_{\tilde b_2},m_{\tilde b_1},m_{\tilde b_2}) \cos^2\theta_{\tilde b}\nonumber \\
& +& \frac{1}{8\pi^2} \frac{g_{\phi22} g_{\phi12}}{g_{\phi bb}^2} t D_{123}(-p_1,p_3,p_4, m_{\tilde g}, m_{\tilde b_2},m_{\tilde b_2},m_{\tilde b_1}) \sin\theta_{\tilde b} \cos\theta_{\tilde b} \nonumber \\
& +& \frac{1}{8\pi^2} \frac{g_{\phi22}^2}{g_{\phi bb}^2} t D_{123}(-p_1,p_3,p_4, m_{\tilde g}, m_{\tilde b_2},m_{\tilde b_2},m_{\tilde b_2}) \cos^2\theta_{\tilde b}\nonumber \\
X_{73} &= & -\frac{1}{8\pi^2} g^2_{\phi11} m_{\tilde g} D_{0}(-p_1,p_3,p_4, m_{\tilde g}, m_{\tilde b_1},m_{\tilde b_1},m_{\tilde b_1}) \sin\theta_{\tilde b}\cos\theta_{\tilde b}\nonumber \\
&-& \frac{1}{8\pi^2} g_{\phi11} g_{\phi12} m_{\tilde g} D_{0}(-p_1,p_3,p_4, m_{\tilde g}, m_{\tilde b_1},m_{\tilde b_1},m_{\tilde b_2}) \cos^2\theta_{\tilde b} \nonumber \\
& -& \frac{1}{8\pi^2} g^2_{\phi12} m_{\tilde g} D_{0}(-p_1,p_3,p_4, m_{\tilde g}, m_{\tilde b_1},m_{\tilde b_2},m_{\tilde b_1})  \sin\theta_{\tilde b}\cos\theta_{\tilde b}\nonumber \\
&-& \frac{1}{8\pi^2} g_{\phi22} g_{\phi12} m_{\tilde g} D_{0}(-p_1,p_3,p_4, m_{\tilde g}, m_{\tilde b_1},m_{\tilde b_2},m_{\tilde b_2}) \cos^2\theta_{\tilde b} \nonumber \\
&+& \frac{1}{8\pi^2} g_{\phi11} g_{\phi12} m_{\tilde g} D_{0}(-p_1,p_3,p_4, m_{\tilde g}, m_{\tilde b_2},m_{\tilde b_1},m_{\tilde b_1}) \sin^2\theta_{\tilde b} \nonumber \\
&+& \frac{1}{8\pi^2} g^2_{\phi12} m_{\tilde g} D_{0}(-p_1,p_3,p_4, m_{\tilde g}, m_{\tilde b_2},m_{\tilde b_1},m_{\tilde b_2}) \sin\theta_{\tilde b}\cos\theta_{\tilde b}\nonumber \\
&+& \frac{1}{8\pi^2} g_{\phi22} g_{\phi12} m_{\tilde g} D_{0}(-p_1,p_3,p_4, m_{\tilde g}, m_{\tilde b_2},m_{\tilde b_2},m_{\tilde b_1}) \sin^2\theta_{\tilde b} \nonumber \\
&+& \frac{1}{8\pi^2} g^2_{\phi22} m_{\tilde g} D_{0}(-p_1,p_3,p_4, m_{\tilde g}, m_{\tilde b_2},m_{\tilde b_2},m_{\tilde b_2}) \sin\theta_{\tilde b}\cos\theta_{\tilde b}\nonumber \\
X_{74} &= & -\frac{1}{8\pi^2} g^2_{\phi11} m_{\tilde g} D_{0}(-p_1,p_3,p_4, m_{\tilde g}, m_{\tilde b_1},m_{\tilde b_1},m_{\tilde b_1}) \sin\theta_{\tilde b}\cos\theta_{\tilde b}\nonumber \\
&+& \frac{1}{8\pi^2} g_{\phi11} g_{\phi12} m_{\tilde g} D_{0}(-p_1,p_3,p_4, m_{\tilde g}, m_{\tilde b_1},m_{\tilde b_1},m_{\tilde b_2}) \sin^2\theta_{\tilde b} \nonumber \\
&-& \frac{1}{8\pi^2} g^2_{\phi12} m_{\tilde g} D_{0}(-p_1,p_3,p_4, m_{\tilde g}, m_{\tilde b_1},m_{\tilde b_2},m_{\tilde b_1})  \sin\theta_{\tilde b}\cos\theta_{\tilde b}\nonumber \\
 &+& \frac{1}{8\pi^2} g_{\phi22} g_{\phi12} m_{\tilde g} D_{0}(-p_1,p_3,p_4, m_{\tilde g}, m_{\tilde b_1},m_{\tilde b_2},m_{\tilde b_2}) \sin^2\theta_{\tilde b} \nonumber \\
 &- & \frac{1}{8\pi^2} g_{\phi11} g_{\phi12} m_{\tilde g} D_{0}(-p_1,p_3,p_4, m_{\tilde g}, m_{\tilde b_2},m_{\tilde b_1},m_{\tilde b_1}) \cos^2\theta_{\tilde b} \nonumber \\
&+& \frac{1}{8\pi^2} g^2_{\phi12} m_{\tilde g} D_{0}(-p_1,p_3,p_4, m_{\tilde g}, m_{\tilde b_2},m_{\tilde b_1},m_{\tilde b_2}) \sin\theta_{\tilde b}\cos\theta_{\tilde b}\nonumber \\
 &- & \frac{1}{8\pi^2} g_{\phi22} g_{\phi12} m_{\tilde g} D_{0}(-p_1,p_3,p_4, m_{\tilde g}, m_{\tilde b_2},m_{\tilde b_2},m_{\tilde b_1}) \cos^2\theta_{\tilde b} \nonumber \\
&+& \frac{1}{8\pi^2} g^2_{\phi22} m_{\tilde g} D_{0}(-p_1,p_3,p_4, m_{\tilde g}, m_{\tilde b_2},m_{\tilde b_2},m_{\tilde b_2}) \sin\theta_{\tilde b}\cos\theta_{\tilde b}\nonumber \\
 \nonumber \\
X_{8i} &= & X_{7i}( t \leftrightarrow u, p_3 \leftrightarrow p_4) \quad i = 1,2,3,4
 \nonumber \\ \nonumber \\
X_{93} &= & -\frac{1}{8\pi^2} g_{h11}g_{h\phi\phi}\frac{ m_{\tilde g}}{(s-M_h^2) + i M_h \Gamma_h} C_{0}(-p_1,p_1+p_2, m_{\tilde g}, m_{\tilde b_1},m_{\tilde b_1}) \sin\theta_{\tilde b} \cos\theta_{\tilde b}\nonumber \\
&-& \frac{1}{8\pi^2} g_{h12}g_{h\phi\phi}\frac{ m_{\tilde g}}{(s-M_h^2) + i M_h \Gamma_h} C_{0}(-p_1,p_1+p_2, m_{\tilde g}, m_{\tilde b_1},m_{\tilde b_2}) \cos^2\theta_{\tilde b}\nonumber \\
& +& \frac{1}{8\pi^2} g_{h12}g_{h\phi\phi}\frac{ m_{\tilde g}}{(s-M_h^2) + i M_h \Gamma_h} C_{0}(-p_1,p_1+p_2, m_{\tilde g}, m_{\tilde b_2},m_{\tilde b_1}) \sin^2\theta_{\tilde b}\nonumber \\
&+& \frac{1}{8\pi^2} g_{h22}g_{h\phi\phi}\frac{ m_{\tilde g}}{(s-M_h^2) + i M_h \Gamma_h} C_{0}(-p_1,p_1+p_2, m_{\tilde g}, m_{\tilde b_2},m_{\tilde b_2}) \sin\theta_{\tilde b} \cos\theta_{\tilde b}\nonumber \\
X_{94} &= & -\frac{1}{8\pi^2} g_{h11}g_{h\phi\phi}\frac{ m_{\tilde g}}{(s-M_h^2) + i M_h \Gamma_h} C_{0}(-p_1,p_1+p_2, m_{\tilde g}, m_{\tilde b_1},m_{\tilde b_1}) \sin\theta_{\tilde b} \cos\theta_{\tilde b}\nonumber \\
&+& \frac{1}{8\pi^2} g_{h12}g_{h\phi\phi}\frac{ m_{\tilde g}}{(s-M_h^2) + i M_h \Gamma_h} C_{0}(-p_1,p_1+p_2, m_{\tilde g}, m_{\tilde b_1},m_{\tilde b_2}) \sin^2\theta_{\tilde b} \nonumber \\
 &- & \frac{1}{8\pi^2} g_{h12}g_{h\phi\phi}\frac{ m_{\tilde g}}{(s-M_h^2) + i M_h \Gamma_h} C_{0}(-p_1,p_1+p_2, m_{\tilde g}, m_{\tilde b_2},m_{\tilde b_1}) \cos^2\theta_{\tilde b} \nonumber \\
 &+& \frac{1}{8\pi^2} g_{h22}g_{h\phi\phi}\frac{ m_{\tilde g}}{(s-M_h^2) + i M_h \Gamma_h} C_{0}(-p_1,p_1+p_2, m_{\tilde g}, m_{\tilde b_2},m_{\tilde b_2}) \sin\theta_{\tilde b} \cos\theta_{\tilde b}\nonumber \\
 \nonumber \\
X_{10i} & = &  X_{9i} ( g_{hjk} \leftrightarrow g_{Hjk}, g_{h\phi\phi} \leftrightarrow g_{H\phi\phi}) \quad i=3, 4; \quad j, k = 1, 2 \nonumber \\
 \nonumber \\
\end{eqnarray}

\subsection{X coefficients for $b \bar{b} \to A A$}
\begin{eqnarray}
X_{11} & = & \frac{1}{8\pi^2} \left[ B_1(-p_1+p_3, m_{\tilde{g}}, m_{\tilde b_1}) \sin^2\theta_{\tilde b} + B_1(-p_1+p_3, m_{\tilde{g}}, m_{\tilde b_2}) \cos^2\theta_{\tilde b}\right] \nonumber \\
X_{12} & = & \frac{1}{8\pi^2} \left[ B_1(-p_1+p_3, m_{\tilde{g}}, m_{\tilde b_1}) \cos^2\theta_{\tilde b} + B_1(-p_1+p_3, m_{\tilde{g}}, m_{\tilde b_2}) \sin^2\theta_{\tilde b}\right] \nonumber \\
X_{13} & = & \frac{1}{8\pi^2} \frac{g_{hbb}^2}{t} m_{\tilde g} \left[B_{0}(-p_1+p_3, m_{\tilde g}, m_{\tilde b_1}) - B_{0}(-p_1+p_3, m_{\tilde g}, m_{\tilde b_2}) \right]\sin\theta_{\tilde b} \cos\theta_{\tilde b} \nonumber \\
X_{14} &=& X_{13} \nonumber \\
 \nonumber \\
X_{2i} &= & X_{1i}( t \leftrightarrow u, p_3 \leftrightarrow p_4) \quad i = 1,2,3,4 \nonumber \\
 \nonumber \\
X_{31} &= &  \frac{1}{8\pi^2}\frac{g_{A12}} {g_{Abb}} m_{\tilde g} C_0(-p_1,p_3, m_{\tilde g}, m_{\tilde b_1},m_{\tilde b_2}) \cos^2\theta_{\tilde b}  \nonumber \\
&+& \frac{1}{8\pi^2}\frac{g_{A12}} {g_{Abb}} m_{\tilde g} C_0(-p_1,p_3, m_{\tilde g}, m_{\tilde b_2},m_{\tilde b_1}) \sin^2\theta_{\tilde b}  \nonumber \\
X_{32} & = & \frac{1}{8\pi^2}\frac{g_{A12}} {g_{Abb}} m_{\tilde g} C_0(-p_1,p_3, m_{\tilde g}, m_{\tilde b_1},m_{\tilde b_2}) \sin^2\theta_{\tilde b}  \nonumber \\
&+& \frac{1}{8\pi^2}\frac{g_{A12}} {g_{Abb}} m_{\tilde g} C_0(-p_1,p_3, m_{\tilde g}, m_{\tilde b_2},m_{\tilde b_1}) \cos^2\theta_{\tilde b}  \nonumber \\
X_{33} &= & -\frac{1}{8\pi^2} g_{Abb} g_{A12} C_{12}(-p_1,p_3, m_{\tilde g}, m_{\tilde b_1},m_{\tilde b_2}) \sin\theta_{\tilde b} \cos\theta_{\tilde b}  \nonumber \\
&+& \frac{1}{8\pi^2} g_{Abb} g_{A12} C_{12}(-p_1,p_3, m_{\tilde g}, m_{\tilde b_2},m_{\tilde b_1}) \sin\theta_{\tilde b} \cos\theta_{\tilde b}  \nonumber \\
X_{34} &= & -\frac{1}{8\pi^2} g_{Abb} g_{A12} C_{12}(-p_1,p_3, m_{\tilde g}, m_{\tilde b_1},m_{\tilde b_2}) \sin\theta_{\tilde b} \cos\theta_{\tilde b}  \nonumber \\
&+& \frac{1}{8\pi^2} g_{Abb} g_{A12} C_{12}(-p_1,p_3, m_{\tilde g}, m_{\tilde b_2},m_{\tilde b_1}) \sin\theta_{\tilde b} \cos\theta_{\tilde b}  \nonumber \\
 \nonumber \\
X_{4i} &= & X_{3i}( t \leftrightarrow u, p_3 \leftrightarrow p_4) \quad i = 1,2,3,4
 \nonumber \\ \nonumber \\
X_{51} &= & \frac{1}{8\pi^2}\frac{g_{A12}} {g_{Abb}} m_{\tilde g} C_0(p_2,-p_4, m_{\tilde g}, m_{\tilde b_2},m_{\tilde b_1}) \sin^2\theta_{\tilde b}  \nonumber \\
&+& \frac{1}{8\pi^2}\frac{g_{A12}} {g_{Abb}} m_{\tilde g} C_0(p_2,-p_4, m_{\tilde g}, m_{\tilde b_1},m_{\tilde b_2}) \cos^2\theta_{\tilde b}  \nonumber \\
X_{52} & = & \frac{1}{8\pi^2}\frac{g_{A12}} {g_{Abb}} m_{\tilde g} C_0(p_2,-p_4, m_{\tilde g}, m_{\tilde b_2},m_{\tilde b_1}) \cos^2\theta_{\tilde b}  \nonumber \\
&+& \frac{1}{8\pi^2}\frac{g_{A12}} {g_{Abb}} m_{\tilde g} C_0(p_2,-p_4, m_{\tilde g}, m_{\tilde b_1},m_{\tilde b_2}) \sin^2\theta_{\tilde b}  \nonumber \\
X_{53} &= & \frac{1}{8\pi^2} g_{Abb} g_{A12} C_{12}(p_2,-p_4, m_{\tilde g}, m_{\tilde b_2},m_{\tilde b_1}) \sin\theta_{\tilde b} \cos\theta_{\tilde b}  \nonumber \\
&-& \frac{1}{8\pi^2} g_{Abb} g_{A12} C_{12}(p_2,-p_4, m_{\tilde g}, m_{\tilde b_1},m_{\tilde b_2}) \sin\theta_{\tilde b} \cos\theta_{\tilde b}  \nonumber \\
X_{54} &= & \frac{1}{8\pi^2} g_{Abb} g_{A12} C_{12}(p_2,-p_4, m_{\tilde g}, m_{\tilde b_2},m_{\tilde b_1}) \sin\theta_{\tilde b} \cos\theta_{\tilde b}  \nonumber \\
&-& \frac{1}{8\pi^2} g_{Abb} g_{A12} C_{12}(p_2,-p_4, m_{\tilde g}, m_{\tilde b_1},m_{\tilde b_2}) \sin\theta_{\tilde b} \cos\theta_{\tilde b}  \nonumber \\
 \nonumber \\
X_{6i} &= & X_{5i}( t \leftrightarrow u, p_3 \leftrightarrow p_4) \quad i = 1,2,3,4
 \nonumber \\ \nonumber \\
X_{71} &=& \frac{1}{8\pi^2} \frac{g_{A12}^2}{g_{Abb}^2} t D_{123}(-p_1,p_3,p_4, m_{\tilde g}, m_{\tilde b_1},m_{\tilde b_2},m_{\tilde b_1}) \cos^2\theta_{\tilde b} \nonumber \\
& +& \frac{1}{8\pi^2} \frac{g_{A12}^2}{g_{Abb}^2} t D_{123}(-p_1,p_3,p_4, m_{\tilde g}, m_{\tilde b_2},m_{\tilde b_1},m_{\tilde b_2}) \sin^2\theta_{\tilde b} \nonumber \\
X_{72} &=& \frac{1}{8\pi^2} \frac{g_{A12}^2}{g_{Abb}^2} t D_{123}(-p_1,p_3,p_4, m_{\tilde g}, m_{\tilde b_1},m_{\tilde b_2},m_{\tilde b_1}) \sin^2\theta_{\tilde b} \nonumber \\
&+& \frac{1}{8\pi^2} \frac{g_{A12}^2}{g_{Abb}^2} t D_{123}(-p_1,p_3,p_4, m_{\tilde g}, m_{\tilde b_2},m_{\tilde b_1},m_{\tilde b_2}) \cos^2\theta_{\tilde b} \nonumber \\
X_{73} &= &  -\frac{1}{8\pi^2} g^2_{A12} m_{\tilde g} D_{0}(-p_1,p_3,p_4, m_{\tilde g}, m_{\tilde b_1},m_{\tilde b_2},m_{\tilde b_1})  \sin\theta_{\tilde b}\cos\theta_{\tilde b}\nonumber \\
&+& \frac{1}{8\pi^2} g^2_{A12} m_{\tilde g} D_{0}(-p_1,p_3,p_4, m_{\tilde g}, m_{\tilde b_2},m_{\tilde b_1},m_{\tilde b_2}) \sin\theta_{\tilde b}\cos\theta_{\tilde b}\nonumber \\
X_{74} &= & -\frac{1}{8\pi^2} g^2_{A12} m_{\tilde g} D_{0}(-p_1,p_3,p_4, m_{\tilde g}, m_{\tilde b_1},m_{\tilde b_2},m_{\tilde b_1})  \sin\theta_{\tilde b}\cos\theta_{\tilde b}\nonumber \\
&+& \frac{1}{8\pi^2} g^2_{A12} m_{\tilde g} D_{0}(-p_1,p_3,p_4, m_{\tilde g}, m_{\tilde b_2},m_{\tilde b_1},m_{\tilde b_2}) \sin\theta_{\tilde b}\cos\theta_{\tilde b}\nonumber \\
 \nonumber \\
X_{8i} &= & X_{7i}( t \leftrightarrow u, p_3 \leftrightarrow p_4) \quad i = 1,2,3,4
 \nonumber \\ \nonumber \\
X_{93} &= & -\frac{1}{8\pi^2} g_{h11}g_{hAA}\frac{ m_{\tilde g}}{(s-M_h^2) + i M_h \Gamma_h} C_{0}(-p_1,p_1+p_2, m_{\tilde g}, m_{\tilde b_1},m_{\tilde b_1}) \sin\theta_{\tilde b} \cos\theta_{\tilde b}\nonumber \\
&-& \frac{1}{8\pi^2} g_{h12}g_{hAA}\frac{ m_{\tilde g}}{(s-M_h^2) + i M_h \Gamma_h} C_{0}(-p_1,p_1+p_2, m_{\tilde g}, m_{\tilde b_1},m_{\tilde b_2}) \cos^2\theta_{\tilde b}\nonumber \\
& +& \frac{1}{8\pi^2} g_{h12}g_{hAA}\frac{ m_{\tilde g}}{(s-M_h^2) + i M_h \Gamma_h} C_{0}(-p_1,p_1+p_2, m_{\tilde g}, m_{\tilde b_2},m_{\tilde b_1}) \sin^2\theta_{\tilde b}\nonumber \\
&+& \frac{1}{8\pi^2} g_{h22}g_{hAA}\frac{ m_{\tilde g}}{(s-M_h^2) + i M_h \Gamma_h} C_{0}(-p_1,p_1+p_2, m_{\tilde g}, m_{\tilde b_2},m_{\tilde b_2}) \sin\theta_{\tilde b} \cos\theta_{\tilde b}\nonumber \\
X_{94} &= & -\frac{1}{8\pi^2} g_{h11}g_{hAA}\frac{ m_{\tilde g}}{(s-M_h^2) + i M_h \Gamma_h} C_{0}(-p_1,p_1+p_2, m_{\tilde g}, m_{\tilde b_1},m_{\tilde b_1}) \sin\theta_{\tilde b} \cos\theta_{\tilde b}\nonumber \\
&+& \frac{1}{8\pi^2} g_{h12}g_{hAA}\frac{ m_{\tilde g}}{(s-M_h^2) + i M_h \Gamma_h} C_{0}(-p_1,p_1+p_2, m_{\tilde g}, m_{\tilde b_1},m_{\tilde b_2}) \sin^2\theta_{\tilde b} \nonumber \\
 &- & \frac{1}{8\pi^2} g_{h12}g_{hAA}\frac{ m_{\tilde g}}{(s-M_h^2) + i M_h \Gamma_h} C_{0}(-p_1,p_1+p_2, m_{\tilde g}, m_{\tilde b_2},m_{\tilde b_1}) \cos^2\theta_{\tilde b} \nonumber \\
 &+& \frac{1}{8\pi^2} g_{h22}g_{hAA}\frac{ m_{\tilde g}}{(s-M_h^2) + i M_h \Gamma_h} C_{0}(-p_1,p_1+p_2, m_{\tilde g}, m_{\tilde b_2},m_{\tilde b_2}) \sin\theta_{\tilde b} \cos\theta_{\tilde b}\nonumber \\
 \nonumber \\
X_{10i} & = &  X_{9i} ( g_{hjk} \leftrightarrow g_{Hjk}, g_{hAA} \leftrightarrow g_{HAA}) \quad i=3, 4; \quad j, k = 1, 2 \nonumber \\
 \nonumber \\
\end{eqnarray}

\newpage
\section{Higgs Couplings in MSSM}

\begin{table}[ht]
\begin{center}
\begin{tabular}{|c|c|c|}
\hline
$g_{hbb}$   & $g_{Hbb}$   & $g_{Abb}$   \cr\hline
 \hspace{1.5cm}$ \frac{g m_b}{2M_W} \frac{\sin\alpha}{\cos\beta}$ 
\hspace{1.5cm} & \hspace{1.5cm} $- \frac{g m_b}{2M_W} \frac{\cos\alpha}{\cos\beta}$ \hspace{1.5cm} & \hspace{1.5cm} $-  \frac{g m_b}{2M_W} \tan\beta$ \hspace{1.5cm} \cr\hline
\end{tabular}
\smallskip
\caption{The $\phi b\bar{b}$ vertex couplings.  For $\phi = h,H$ the Feynman rule is $ig_{\phi bb}$ and
for $\phi=A$ the Feynman rule is $\gamma_5 g_{Abb}$.}
\label{coupling:LO}
\end{center}
\end{table}
\begin{table}[htb]
\begin{center}
\begin{tabular}{|c|c|c|}
\hline
$g$   & $h$   & $H$   \cr\hline
$hh$ & $-3 \frac{gM_Z}{2\cos\theta_W} \cos2\alpha \sin(\beta+\alpha) $ & $- \frac{gM_Z}{2\cos\theta_W} [2\sin2\alpha \sin(\beta+\alpha)-\cos(\beta+\alpha)\cos2\alpha] $  \cr\hline
$HH$ & $ \frac{g M_Z}{2\cos\theta_W} [2 \sin2\alpha \cos(\beta+\alpha)+
\sin(\beta+\alpha)\cos2\alpha] $  &  $-3 \frac{g M_Z}{2 \cos\theta_W} \cos2\alpha \cos(\beta+\alpha) $  \cr\hline
$AA$ & $-\frac{g M_Z}{2\cos\theta_W} \cos2\beta \sin(\beta+\alpha) $  &  $  \frac{g M_Z}{2\cos\theta_W}\cos2\beta \cos(\beta+\alpha) $  \cr\hline
\end{tabular}
\smallskip
\caption{The  $\phi_i\phi_j\phi_j$ triple Higgs vertices. 
For $\phi = h,H,A$ the Feynman rule is $ig_{\phi_i\phi_j\phi_j}$.}
\label{coupling:ghhh}
\end{center}
\end{table}
\begin{table}[htb]
%\begin{flushleft}
\begin{tabular}{|c|c|}
\hline
$g_{h11}$ & \hspace{.5cm}$ \frac{g m_b}{2M_W \cos\beta} [\sin2\theta_b(\mu \cos\alpha +A_b \sin\alpha)] 
+  \frac{g m_b^2}{M_W \cos\beta} \sin\alpha $ \cr
&\cr
 & $- \frac{g M_W}{2} \sin(\beta+\alpha)[(1-\frac{\tan^2\theta_w}{3})\cos^2\theta_b + \frac{2}{3} \tan^2\theta_W] $ \cr\hline
$g_{h22}$ & \hspace{.5cm}$ \frac{g m_b}{2M_W \cos\beta} [-\sin2\theta_b(\mu \cos\alpha +A_b \sin\alpha)] 
+  \frac{g m_b^2}{M_W \cos\beta} \sin\alpha $ \cr
&\cr
 & $- \frac{g M_W}{2} \sin(\beta+\alpha)[(1-\frac{\tan^2\theta_w}{3})\sin^2\theta_b + \frac{2}{3} \tan^2\theta_W] $ \cr\hline
$g_{h12}$ & \hspace{.5cm}$ \frac{g m_b}{2M_W \cos\beta} \cos2\theta_b(\mu \cos\alpha +A_b \sin\alpha)] $ \cr
&\cr
 & $ +\frac{g M_W}{4} \sin2\theta_b \sin(\beta+\alpha)(1-\frac{\tan^2\theta_w}{3}) $ \cr\hline\hline
$g_{H11}$ & \hspace{.5cm}$ \frac{g m_b}{2M_W \cos\beta} (\sin2\theta_b)(\mu \sin\alpha - A_b \cos\alpha)] -  \frac{g m_b^2}{M_W \cos\beta} \cos\alpha $ \cr
&\cr 
& $+ \frac{g M_W}{2} \cos(\beta+\alpha)[(1-\frac{\tan^2\theta_w}{3})\cos^2\theta_b + \frac{2}{3} \tan^2\theta_W] $ \cr\hline
$g_{H22}$ & \hspace{.5cm}$ -\frac{g m_b}{2M_W \cos\beta} \sin2\theta_b(\mu \sin\alpha -A_b \cos\alpha) -  \frac{g m_b^2}{M_W \cos\beta} \cos\alpha $ \cr
&\cr
 & $+ \frac{g M_W}{2} \cos(\beta+\alpha)[(1-\frac{\tan^2\theta_w}{3})\sin^2\theta_b + \frac{2}{3} \tan^2\theta_W] $ \cr\hline
$g_{H12}$ & \hspace{.5cm}$ \frac{g m_b}{2M_W \cos\beta} \cos2\theta_b(\mu \sin\alpha -A_b \cos\alpha) $ \cr
&\cr
 & $- \frac{g M_W}{4} \sin2\theta_b \cos(\beta+\alpha)(1-\frac{\tan^2\theta_w}{3}) $ \cr\hline\hline
$g_{A12}$ & \hspace{.5cm}$ \frac{g m_b}{2M_W \cos\beta} (\mu \cos\beta + A_b \sin\beta) $ \cr \hline
$g_{A21}$ & $-g_{A12}$  \cr\hline
\end{tabular}
\smallskip
\caption{Higgs-sbottom-sbottom couplings. For $\phi = h,H$ the Feynman rule is $ig_{\phi {\tilde b_i}{\tilde b_j}}$
and for $\phi=A$ the Feynman rule is $g_{A{\tilde b_i}{\tilde b_j}}$.}
\label{coupling:gs}
%\end{left}
\end{table}

\newpage
%-----------------------------------------------------------------------
% Bibliography
%-----------------------------------------------------------------------

%-----------------------------------------------------------------------
%   END DOCUMENT
%-----------------------------------------------------------------------

\begin{thebibliography}{20}
%----------------------------------------------
% SUSY
%----------------------------------------------
%\cite{Nilles:1983ge}
\bibitem{Nilles:1983ge}
  H.~P.~Nilles,
  %``Supersymmetry, Supergravity And Particle Physics,''
  Phys.\ Rept.\  {\bf 110}, 1 (1984);
  %%CITATION = PRPLC,110,1;%%
%\cite{Gunion:1989we}
\bibitem{Gunion:1989we}
  J.~F.~Gunion, H.~E.~Haber, G.~L.~Kane and S.~Dawson,
   {\it The Higgs Hunter's Guide} (Addison-Wesley, Menlo Park, 1990).
%SCIPP-89/13
%\href{http://www.slac.stanford.edu/spires/find/hep/www?r=scipp-89\%2F13}{SPIRES entry}
%------------------------------------------------
% LO to gg -> hh
%------------------------------------------------
%\cite{Dicus:1987ic}
\bibitem{Dicus:1987ic}
  D.~A.~Dicus, C.~Kao and S.~S.~D.~Willenbrock,
  %``HIGGS BOSON PAIR PRODUCTION FROM GLUON FUSION,''
  Phys.\ Lett.\ B {\bf 203}, 457 (1988).
  %%CITATION = PHLTA,B203,457;%%
%\cite{Glover:1987nx}
\bibitem{Glover:1987nx}
  E.~W.~N.~Glover and J.~J.~van der Bij,
  %``HIGGS BOSON PAIR PRODUCTION VIA GLUON FUSION,''
  Nucl.\ Phys.\ B {\bf 309}, 282 (1988).
  %%CITATION = NUPHA,B309,282;%%
%------------------------------------------------
% NLO to gg -> hh
%------------------------------------------------
%\cite{Plehn:1996wb}
\bibitem{Plehn:1996wb}
  T.~Plehn, M.~Spira and P.~M.~Zerwas,
  %``Pair Production of Neutral Higgs Particles in Gluon--Gluon Collisions,''
  Nucl.\ Phys.\  B {\bf 479}, 46 (1996)
  [Erratum-ibid.\  B {\bf 531}, 655 (1998)]
  [arXiv:hep-ph/9603205].
  %%CITATION = NUPHA,B479,46;%%

%\cite{Dawson:1998py}
\bibitem{Dawson:1998py}
  S.~Dawson, S.~Dittmaier and M.~Spira,
  %``Neutral Higgs-boson pair production at hadron colliders:  {QCD}
  %  corrections,''
  Phys.\ Rev.\ D {\bf 58}, 115012 (1998)
  [arXiv:hep-ph/9805244].
  %%CITATION = HEP-PH 9805244;%%
%
%\cite{Belyaev:1999mx}
\bibitem{Belyaev:1999mx}
  A.~Belyaev, M.~Drees, O.~J.~P.~Eboli, J.~K.~Mizukoshi and S.~F.~Novaes,
  %``Supersymmetric Higgs pair production at hadron colliders,''
  Phys.\ Rev.\  D {\bf 60}, 075008 (1999)
  [arXiv:hep-ph/9905266].
  %%CITATION = PHRVA,D60,075008;%%


%\cite{BarrientosBendezu:2001di}
\bibitem{BarrientosBendezu:2001di}
  A.~A.~Barrientos Bendezu and B.~A.~Kniehl,
  %``Pair production of neutral Higgs bosons at the CERN Large Hadron
  %  Collider,''
  Phys.\ Rev.\ D {\bf 64}, 035006 (2001)
  [arXiv:hep-ph/0103018].
  %%CITATION = HEP-PH 0103018;%% SUSY
%\cite{Binoth:2006ym}
\bibitem{Binoth:2006ym}
  T.~Binoth, S.~Karg, N.~Kauer and R.~Ruckl,
  %``Multi-Higgs boson production in the standard model and beyond,''
  Phys.\ Rev.\  D {\bf 74}, 113008 (2006)
  [arXiv:hep-ph/0608057].
  %%CITATION = PHRVA,D74,113008;%%
%-------------------------------------------------------
% NLO bb->hh in SM
%-------------------------------------------------------
%\cite{Dawson:2006dm}
\bibitem{Dawson:2006dm}
  S.~Dawson, C.~Kao, Y.~Wang and P.~Williams,
  %``QCD corrections to Higgs pair production in bottom quark fusion,''
  Phys.\ Rev.\  D {\bf 75}, 013007 (2007)  
  [arXiv:hep-ph/0610284].
  %%CITATION = HEP-PH 0610284;%%
%-------------------------------------------------------
%NLO bb->H_i H_j in MSSM
%-------------------------------------------------------
%\cite{Jin:2005gw}
\bibitem{Jin:2005gw}
  L.~G.~Jin, C.~S.~Li, Q.~Li, J.~J.~Liu and R.~J.~Oakes,
  %``Next-to-leading order QCD predictions for pair production of neutral  Higgs
  %bosons at the CERN Large Hadron Collider,''
  Phys.\ Rev.\  D {\bf 71}, 095004 (2005)
  [arXiv:hep-ph/0501279].
  %%CITATION = PHRVA,D71,095004;%%
%------------------------------------------------
% Trilinear Higgs coupling
%------------------------------------------------
%\cite{Boudjema:1995cb}
\bibitem{Boudjema:1995cb}
  F.~Boudjema and E.~Chopin,
  %``Double Higgs production at the linear colliders and the probing of the
  %Higgs selfcoupling,''
  Z.\ Phys.\  C {\bf 73}, 85 (1996)
  [arXiv:hep-ph/9507396].
  %%CITATION = ZEPYA,C73,85;%%

\bibitem{Djouadi:1999rc}
  A.~Djouadi, W.~Kilian, M.~Muhlleitner and P.~M.~Zerwas,
  %``Production of neutral Higgs-boson pairs at LHC,''
  Eur.\ Phys.\ J.\ C {\bf 10}, 45 (1999)
  [arXiv:hep-ph/9904287].
  %%CITATION = HEP-PH 9904287;%%

\bibitem{Muhlleitner:2003me}
M.~Muhlleitner and M.~Spira,
Phys.\ Rev.\ D {\bf 68}, 117701 (2003).

\bibitem{Baur:2003gp}
  U.~Baur, T.~Plehn and D.~L.~Rainwater,
  %``Probing the Higgs self-coupling at hadron colliders using rare decays,''
  Phys.\ Rev.\ D {\bf 69}, 053004 (2004)
  [arXiv:hep-ph/0310056].
  %%CITATION = HEP-PH 0310056;%%

\bibitem{Moretti:2004wa}
  M.~Moretti, S.~Moretti, F.~Piccinini, R.~Pittau and A.~D.~Polosa,
%   ``Higgs boson self-couplings at the LHC as a probe of extended Higgs
%  %sectors,''
  JHEP {\bf 0502}, 024 (2005)
  [arXiv:hep-ph/0410334].
  %%CITATION = HEP-PH 0410334;%%
%----
\bibitem{Maltoni:2003pn}
     Maltoni, F. and Sullivan, Z. and Willenbrock, S.,
     Phys.\  Rev. \ D {\bf 67}, 093005 (2003)
     [arXiv:hep-ph/0301033].

% bb-> h

%\cite{Harlander:2003ai}
\bibitem{Harlander:2003ai}
  R.~V.~Harlander and W.~B.~Kilgore,
%   ``Higgs boson production in bottom quark fusion at  next-to-next-to-leading
%  %order,''
  Phys.\ Rev.\  D {\bf 68}, 013001 (2003)
  [arXiv:hep-ph/0304035].
  %%CITATION = PHRVA,D68,013001;%%

%\cite{Dittmaier:2006cz}
\bibitem{Dittmaier:2006cz}
  S.~Dittmaier, M.~Kramer, A.~Muck and T.~Schluter,
  %``MSSM Higgs-boson production in bottom-quark fusion: Electroweak radiative
  %corrections,''
  JHEP {\bf 0703}, 114 (2007)
  [arXiv:hep-ph/0611353].
  %%CITATION = JHEPA,0703,114;%%


% bg-> bh
%\cite{Campbell:2004pu}
\bibitem{Campbell:2004pu}
  J.~Campbell {\it et al.},
  %``Higgs boson production in association with bottom quarks,''
  [arXiv:hep-ph/0405302].
  %%CITATION = HEP-PH/0405302;%%

%\cite{Dawson:2005vi}
\bibitem{Dawson:2005vi}
  S.~Dawson, C.~B.~Jackson, L.~Reina and D.~Wackeroth,
  %``Higgs production in association with bottom quarks at hadron colliders,''
  Mod.\ Phys.\ Lett.\  A {\bf 21}, 89 (2006)
  [arXiv:hep-ph/0508293].
  %%CITATION = MPLAE,A21,89;%%

%\cite{Dawson:2004wq}
\bibitem{Dawson:2004wq}
  S.~Dawson, C.~B.~Jackson, L.~Reina and D.~Wackeroth,
  %``Higgs boson production with bottom quarks at hadron colliders,''
  Int.\ J.\ Mod.\ Phys.\  A {\bf 20}, 3353 (2005)
  [arXiv:hep-ph/0409345].
  %%CITATION = IMPAE,A20,3353;%%

%\cite{Dawson:2003kb}
\bibitem{Dawson:2003kb}
  S.~Dawson, C.~B.~Jackson, L.~Reina and D.~Wackeroth,
  %``Exclusive Higgs boson production with bottom quarks at hadron  colliders,''
  Phys.\ Rev.\  D {\bf 69}, 074027 (2004)
  [arXiv:hep-ph/0311067].
  %%CITATION = PHRVA,D69,074027;%%

%\cite{Dawson:2004sh}
\bibitem{Dawson:2004sh}
  S.~Dawson, C.~B.~Jackson, L.~Reina and D.~Wackeroth,
  %``Higgs boson production with one bottom quark jet at hadron colliders,''
  Phys.\ Rev.\ Lett.\  {\bf 94}, 031802 (2005)
  [arXiv:hep-ph/0408077].
  %%CITATION = PRLTA,94,031802;%%

%\cite{Dittmaier:2003ej}
\bibitem{Dittmaier:2003ej}
  S.~Dittmaier, M.~Kramer and M.~Spira,
  %``Higgs radiation off bottom quarks at the Tevatron and the LHC,''
  Phys.\ Rev.\  D {\bf 70}, 074010 (2004)
  [arXiv:hep-ph/0309204].
  %%CITATION = PHRVA,D70,074010;%%

%\cite{Dawson:2007ur}
\bibitem{Dawson:2007ur}
  S.~Dawson and C.~B.~Jackson,
  %``SUSY QCD Corrections to Associated Higgs-bottom Quark Production,''
  arXiv:0709.4519 [hep-ph].
  %%CITATION = ARXIV:0709.4519;%%

%------------------------------------------------
% Heavy quark pdfs
%------------------------------------------------
\bibitem{Barnett:1987jw}
R.~M.~Barnett, H.~E.~Haber and D.~E.~Soper,
Nucl.\ Phys.\ B {\bf 306}, 697 (1988).

\bibitem{Olness:1987:ae}
F.~I.~Olness and W.~K.~Tung,
Int.\ J.\ Mod.\ Phys.\ A {\bf 2}, 1413 (1987).

%\cite{Stelzer:1998ni}
\bibitem{Stelzer:1998ni}
  T.~Stelzer, Z.~Sullivan and S.~Willenbrock,
  %``Single top quark production at hadron colliders,''
  Phys.\ Rev.\ D {\bf 58}, 094021 (1998)
  [arXiv:hep-ph/9807340].
  %%CITATION = HEP-PH 9807340;%%
\bibitem{Dicus:1998hs}
  D.~Dicus, T.~Stelzer, Z.~Sullivan and S.~Willenbrock,
  %``Higgs boson production in association with bottom quarks at
  %  next-to-leading order,''
  Phys.\ Rev.\ D {\bf 59}, 094016 (1999)
  [arXiv:hep-ph/9811492].
  %%CITATION = HEP-PH 9811492;%%
 %--------------------------------------------
 %Scalar functions.
 %-------------------------------------------
 %\cite{'tHooft:1972fi}
 \bibitem{'tHooft:1972fi}
   G.~'t Hooft and M.~J.~G.~Veltman,
   %``Regularization And Renormalization Of Gauge Fields,''
   Nucl.\ Phys.\  B {\bf 44}, 189 (1972).
   %%CITATION = NUPHA,B44,189;%%
 %\cite{'tHooft:1978xw}
 \bibitem{'tHooft:1978xw}
   G.~'t Hooft and M.~J.~G.~Veltman,
   %``Scalar One Loop Integrals,''
   Nucl.\ Phys.\  B {\bf 153}, 365 (1979).
   %%CITATION = NUPHA,B153,365;%%
 
 %\cite{van Oldenborgh:1989wn}
 \bibitem{van Oldenborgh:1989wn}
   G.~J.~van Oldenborgh and J.~A.~M.~Vermaseren,
   %``New Algorithms for One Loop Integrals,''
   Z.\ Phys.\  C {\bf 46}, 425 (1990).
   %%CITATION = ZEPYA,C46,425;%%
 
 %\cite{van Oldenborgh:1990yc}
 \bibitem{van Oldenborgh:1990yc}
   G.~J.~van Oldenborgh,
   %``FF: A Package to evaluate one loop Feynman diagrams,''
   Comput.\ Phys.\ Commun.\  {\bf 66}, 1 (1991).
   %%CITATION = CPHCB,66,1;%%
 
%------------------------------------------------
% ACOT
%------------------------------------------------
\bibitem{Aivazis:1993pi}
M.~A.~Aivazis, J.~C.~Collins, F.~I.~Olness and W.~K.~Tung,
%``Leptoproduction of heavy quarks. 2. A Unified QCD formulation of charged
%and neutral current processes from fixed target to collider energies,''
Phys.\ Rev.\ D {\bf 50}, 3102 (1994) [arXiv:hep-ph/9312319].
%%CITATION = HEP-PH 9312319;%%

\bibitem{Collins:1998rz}
J.~C.~Collins,
%``Hard-scattering factorization with heavy quarks: A general treatment,''
Phys.\ Rev.\ D {\bf 58}, 094002 (1998) [arXiv:hep-ph/9806259].
%%CITATION = HEP-PH 9806259;%%

\bibitem{Kramer:2000hn}
M.~Kramer, F.~I.~Olness and D.~E.~Soper,
%``Treatment of heavy quarks in deeply inelastic scattering,''
Phys.\ Rev.\ D {\bf 62}, 096007 (2000) [arXiv:hep-ph/0003035].
%%CITATION = HEP-PH 0003035;%%
%---------------------------------------------------------------
% renormalization
%---------------------------------------------------------------
%\cite{Braaten:1980yq}
\bibitem{Braaten:1980yq}
  E.~Braaten and J.~P.~Leveille,
  %``Higgs Boson Decay And The Running Mass,''
  Phys.\ Rev.\ D {\bf 22}, 715 (1980).
  %%CITATION = PHRVA,D22,715;%%
%\cite{Beenakker:1993yr}
\bibitem{Beenakker:1993yr}
  W.~Beenakker, A.~Denner, W.~Hollik, R.~Mertig, T.~Sack and D.~Wackeroth,
  %``Electroweak one loop contributions to top pair production in hadron
  %colliders,''
  Nucl.\ Phys.\  B {\bf 411}, 343 (1994).
  %%CITATION = NUPHA,B411,343;%%
%\cite{Pierce:1997wu}
\bibitem{Pierce:1997wu}
  D.~M.~Pierce,
  %``Renormalization of supersymmetric theories,''
  arXiv:hep-ph/9805497.
  %%CITATION = HEP-PH/9805497;%%

%\cite{Hafliger:2005aj}
\bibitem{Hafliger:2005aj}
  P.~Hafliger and M.~Spira,
  %``Associated Higgs boson production with heavy quarks in e+ e-  collisions:
  %SUSY-QCD corrections,''
  Nucl.\ Phys.\  B {\bf 719}, 35 (2005)
  [arXiv:hep-ph/0501164].
  %%CITATION = NUPHA,B719,35;%%
%\cite{Berge:2007dz}
\bibitem{Berge:2007dz}
  S.~Berge, W.~Hollik, W.~M.~Mosle and D.~Wackeroth,
   ``SUSY QCD one-loop effects in (un)polarized top-pair production at hadron
  %colliders,''
  Phys.\ Rev.\  D {\bf 76}, 034016 (2007)
  [arXiv:hep-ph/0703016].
  %%CITATION = PHRVA,D76,034016;%%

%------------------------------------------------
%Higher loop correction. mass counterterms.
%-----------------------------------------------
%\cite{Hall:1993gn}
\bibitem{Hall:1993gn}
  L.~J.~Hall, R.~Rattazzi and U.~Sarid,
  %``The Top quark mass in supersymmetric SO(10) unification,''
  Phys.\ Rev.\  D {\bf 50}, 7048 (1994)
  [arXiv:hep-ph/9306309].
  %%CITATION = PHRVA,D50,7048;%%

%\cite{Carena:1999py}
\bibitem{Carena:1999py}
  M.~Carena, D.~Garcia, U.~Nierste and C.~E.~M.~Wagner,
  %``Effective Lagrangian for the anti-t b H+ interaction in the MSSM and
  %charged Higgs phenomenology,''
  Nucl.\ Phys.\  B {\bf 577}, 88 (2000)
  [arXiv:hep-ph/9912516].
  %%CITATION = NUPHA,B577,88;%%

%\cite{Guasch:2003cv}
\bibitem{Guasch:2003cv}
  J.~Guasch, P.~Hafliger and M.~Spira,
  %``MSSM Higgs decays to bottom quark pairs revisited,''
  Phys.\ Rev.\  D {\bf 68}, 115001 (2003)
  [arXiv:hep-ph/0305101].
  %%CITATION = PHRVA,D68,115001;%%

%\cite{Carena:2006ai}
\bibitem{Carena:2006ai}
  M.~S.~Carena, A.~Menon, R.~Noriega-Papaqui, A.~Szynkman and C.~E.~M.~Wagner,
%   ``Constraints on B and Higgs physics in minimal low energy supersymmetric
  %models,''
  Phys.\ Rev.\  D {\bf 74}, 015009 (2006)
  [arXiv:hep-ph/0603106].
  %%CITATION = PHRVA,D74,015009;%%

%------------------------------------------------
% CTEQ PDFs
%------------------------------------------------
%\cite{Pumplin:2002vw}
\bibitem{Pumplin:2002vw}
  J.~Pumplin, D.~R.~Stump, J.~Huston, H.~L.~Lai, P.~Nadolsky and W.~K.~Tung,
  %``New generation of parton distributions with uncertainties from global  QCD
  %  analysis,''
  JHEP {\bf 0207}, 012 (2002)
  [arXiv:hep-ph/0201195].
  %%CITATION = HEP-PH 0201195;%%

%------------------------------------------------
% alpha_s
%------------------------------------------------
%\cite{Marciano:1983pj}
\bibitem{Marciano:1983pj}
  W.~J.~Marciano,
  %``Flavor Thresholds And Lambda In The Modified Minimal Subtraction
  %  Prescription,''
  Phys.\ Rev.\ D {\bf 29}, 580 (1984).
  %%CITATION = PHRVA,D29,580;%%

%\cite{Nason:1987xz}
\bibitem{Nason:1987xz}
  P.~Nason, S.~Dawson and R.~K.~Ellis,
  %``The Total Cross-Section for the Production of Heavy Quarks in Hadronic
  %Collisions,''
  Nucl.\ Phys.\  B {\bf 303}, 607 (1988).
  %%CITATION = NUPHA,B303,607;%%
%------------------------------------------------
% Running mass
%------------------------------------------------
%\cite{Vermaseren:1997fq}
\bibitem{Vermaseren:1997fq}
  J.~A.~M.~Vermaseren, S.~A.~Larin and T.~van Ritbergen,
  %``The 4-loop quark mass anomalous dimension and the invariant quark  mass,''
  Phys.\ Lett.\ B {\bf 405}, 327 (1997)
  [arXiv:hep-ph/9703284].
  %%CITATION = HEP-PH 9703284;%%
%---------------------------------------
% SUSY coupling
%-----------------------------------
%\cite{Haber:1984rc}
\bibitem{Haber:1984rc}
  H.~E.~Haber and G.~L.~Kane,
  %``The Search For Supersymmetry: Probing Physics Beyond The Standard Model,''
  Phys.\ Rept.\  {\bf 117}, 75 (1985).
  %%CITATION = PRPLC,117,75;%%

\bibitem{Drees:2004}
M.~Drees, R.~M.Godbole and P.~Roy, {\it Theory and Phenomenology of Sparticles : An Account of Four-diemnsional N=1 Supersymmetry in High Energy Physics} (World Scientific)

\bibitem{Tata:2006}
  H.~Haer, X.~Tata, {\it
 Weak Scale Supersymmetry : From Superfields to Scattering Events} (Cambridge University Press)

%------------------------------------------
% Higgs Boson mass
%-------------------------------------------
 %\cite{Baer:1991yc}
 \bibitem{Baer:1991yc}
   H.~Baer, M.~Bisset, C.~Kao and X.~Tata,
   %``Observability of gamma gamma decays of Higgs bosons from supersymmetry at
   %hadron supercolliders,''
   Phys.\ Rev.\  D {\bf 46}, 1067 (1992).
   %%CITATION = PHRVA,D46,1067;%%

%------------------------------------------
%decoupling behavior
%----------------------------------------
%\cite{Haber:2000kq}
\bibitem{Haber:2000kq}
  H.~E.~Haber, M.~J.~Herrero, H.~E.~Logan, S.~Penaranda, S.~Rigolin and D.~Temes,
  %``SUSY-QCD corrections to the MSSM h0 b anti-b vertex in the decoupling
  %limit,''
  Phys.\ Rev.\  D {\bf 63}, 055004 (2001)
  [arXiv:hep-ph/0007006].
  %%CITATION = PHRVA,D63,055004;%%

\end{thebibliography}
\end{document}